\newif\ifDAG\DAGtrue
\documentclass[12pt, enabledeprecatedfontcommands]{scrartcl}

\setkomafont{disposition}{\bfseries}

 \usepackage{geometry}

\usepackage{fancyhdr}

\newlength\FHoffset
\fancyheadoffset{\FHoffset}

\newlength\FHleft
\newlength\FHright

\newbox\FHline
\setbox\FHline=\hbox{\hsize=\paperwidth%
  \hspace*{\FHleft}%
  \rule{\dimexpr\headwidth-\FHleft-\FHright\relax}{\headrulewidth}\hspace*{\FHright}%
}

\usepackage{moreverb,url}
\usepackage{placeins}

\renewcommand{\thefootnote}{\Roman{footnote}}

\usepackage{amsmath,amsthm,amssymb,amsfonts}
\usepackage{graphicx, xfrac, MnSymbol}
\usepackage{xcolor}
\usepackage{epstopdf}
\usepackage{dsfont}
\usepackage{booktabs}

\usepackage{tikz}
\usetikzlibrary{shapes.misc}

\usepackage{url}

\usepackage{hyperref}

\usepackage[superscript,biblabel]{cite}

\definecolor{darkgreen}{rgb}{0,0.5,0}
\newcommand{\logit}{\mbox{logit}}
\newcommand{\fvol}{f_{\mbox{vol}}}
\newcommand{\ind}{\stackrel{\mbox{ind.}}{\sim}}
\newcommand{\iid}{\stackrel{\mbox{iid}}{\sim}}

\newcommand{\vol}{v}

\newcommand{\customfootnotetext}[2]{{% Group to localize change to footnote
  \renewcommand{\thefootnote}{#1}% Update footnote counter representation
  \footnotetext[0]{#2}}}
\begin{document}

%\raggedright

\title{Modelling volume-outcome relationships in health care}

\author{\normalsize Maurilio Gutzeit\textsuperscript{$\ast$}, Johannes Rauh, Maximilian K{\"a}hler and Jona Cederbaum\\ \small Federal Institute for Quality Assurance and Transparency in Healthcare, Berlin, Germany}

\maketitle

\customfootnotetext{$\ast$}{Corresponding author. E-mail: maurilio.gutzeit@iqtig.org}

\begin{abstract}
\noindent \small Despite the ongoing strong interest in associations between quality of care and the volume of health care providers, a unified statistical framework for analyzing them is missing, and many studies suffer from poor statistical modelling choices.
We propose a flexible, additive mixed model for studying volume-outcome associations in health care that takes into account individual patient characteristics as well as provider-specific effects through a multi-level approach.
More specifically, we treat volume as a continuous variable, and its effect on the considered outcome is modelled as a smooth function. We take account of different case-mixes by including patient-specific risk factors and of clustering on the provider level through random intercepts. This strategy enables us to extract a smooth volume effect as well as volume-independent provider effects. These two quantities can be compared directly in terms of their magnitude, which gives insight into the sources of variability of quality of care. Based on a causal DAG, we derive conditions under which the volume-effect can be interpreted as a causal effect. The paper provides confidence sets for each of the estimated quantities relying on joint estimation of all effects and parameters. Our approach is illustrated through simulation studies and an application to German health care data.
\end{abstract}

\section{Introduction}\label{sec:intro}

Does the quality of medical care of a health care provider improve with the number of treated patients?
The answer to this question and also the magnitude of such a volume-outcome effect depends strongly on the medical procedure or intervention under consideration and on the considered adverse event, e.g. mortality, complications or readmissions.
Volume-outcome relationships have been found, for instance, for cardiovascular procedures\cite{birkmeyer2002hospital}, pancreatic surgery\cite{krautz2018effect}, treatment of breast cancer\cite{mikeljevic2003surgeon}, orthopaedic procedures such as hip and knee arthroplasties\cite{taylor1997relationship} and care of very low birth weight infants\cite{rogowski2004indirect,Heller2018:Regionalisierung}. 
While the usual expectation is that the number of adverse events is smaller in hospitals with larger volume (decreasing volume-outcome relationship), there are also examples of increasing\cite{LeeLin2007:volout_psych} and non-monotonic volume-outcome relationships\cite{luft1987,GrouvenKuechenhoffSchraederBender08:Review_minimum_provider_volumes,NimptschMansky17:25Volout}.

Findings regarding volume-outcome relationships are important for political and administrative decisions about national and regional health care systems. 
For example, increasing the treatment quality was a major goal when Portugal reformed perinatal care and closed providers with low volume\cite{Neto06:PerinatalPortugal} and
when Denmark decided to reform and centralize its entire hospital system\cite{ChristiansenVrangbaek18:CentralizationDenmark}.
As another example, in the German health care system, threshold values of minimum caseloads regulate specific stationary medical treatments for which there is evidence in favor of a decreasing volume-outcome relationship, such as liver and kidney transplantations or the treatment of very low birth weight infants\cite{MiMeRegelungen2021,hemschemeier2019}.
Understanding the connections between volume and outcome can thus enable sensible administrative decisions and lead to effective interventions or strategies.
Evidence on volume-outcome relationships is also used by private and public healthcare purchasers, such as the CMS or the Leapfrog Group, to selectively refer specific patients to larger providers\cite{CMS:COP_2007,Leapfrog:Scoring2021}.

Several mechanisms have been put forward to explain the decreasing volume-outcome relationship,  which can act simultaneously\cite{luft1987}: On the one hand, higher volume may result in higher quality due to a training effect (\textit{practice makes perfect}). Also, larger hospitals may have more specialist staff and equipment\cite{christian2005}.
According to this explanation, volume can be regarded as a proxy, which loses its importance when the provider properties that influence the outcome more directly, such as experience and equipment, are accessible.
On the other hand, reversing the causal direction, higher quality may lead to higher demand (\textit{selective referral}). These different mechanisms illustrate the delicate question of causality and make clear that it is important to understand where the quality differences between providers come from in order to implement effective interventions.
If, for instance, the presence of special equipment is a main driver for the quality of a treatment, interventions that aim at structural requirements may be more efficient than minimum caseload requirements. However, the intervention on more direct effects is often not possible or associated with a great deal of effort. We provide a causal DAG (directed acyclic graph) which helps formalising causal mechanisms and makes assumed dependencies between the relevant quantities transparent.

From a statistical point of view, the volume outcome problem is a regression problem, where the outcome is regressed on the volume and other variables. The modelling choices that have been employed in prior volume-outcome analyses are quite varied. A unified statistical framework is missing, which makes it difficult to compare results. The different approaches have different advantages, but some also exhibit critical statistical problems. This may lead to a biased estimate of the volume-outcome relationship as well as a bias when estimating significance or statistical uncertainty.

Our approach combines different statistical techniques, avoids known problems, enables valuable conclusions that other approaches do not offer, and performs well for a large variety of settings relevant for volume-outcome relationships. We achieve all this by constructing a flexible generalized additive mixed model (GAMM) which includes all covariates at once in a coherent fashion and allows joint estimation of all effects.

As a central feature of our approach, we treat volume as a continuous variable. In contrast to that, the common practice for volume-outcome analyses consists in discretising volume into a small number of groups\cite{birkmeyer2002hospital, mikeljevic2003surgeon, varagunam2015relationship}. Such discretizations are in general considered as poor statistical practice\cite{royston2006, christian2005}. If volume is treated as a continuous variable at all, its effect is often modelled linearly\cite{NimptschMansky17:25Volout, fischer2017}.
We propose to use a smooth volume effect that allows to model the effect in a more flexible, realistic and less arbitrary way. The functional form and the amount of smoothness is estimated from the data.
Our approach thus allows to identify simple linear as well as nonlinear, monotonous as well as non-monotonous associations. This also enables meaningful comparisons across different volume-outcome studies, which are difficult when different studies discretize volume in different ways.

Furthermore, we take into account the clustered structure of the data which arises from the patients being nested within providers. 
We implement this through random intercepts, which exhibit a number of advantages explained in detail below.
In any case, clustering should not be ignored, since otherwise estimates of the volume effect may be biased and the statistical uncertainty will often be underestimated\cite{rogowski2004indirect,george2017,urbach2005}.

We directly compare the effect of volume on the outcome with that of other provider properties not related to the volume in order to assess how much of the variability of quality of care can be attributed to volume-related differences compared with other differences between providers. This assessment is important, because the variability of the quality of care among hospitals of comparable volume has, for instance, been put forward as an argument against minimum caseload requirements\cite{KutschmannBungardKoettingTruemmerFuschVeit2012:VLBWs}.

We are not aware of previous analyses that offer such a flexible and general approach where all components are jointly modelled allowing for appropriate uncertainty quantification; let alone the additional level of insight through the comparison of provider effects just mentioned. 
Previous comparable publications did not report the estimated volume effect itself or did not offer significance statements and confidence bands, which are crucial for critical reflection of results. 

We illustrate our approach in a simulation study as well as an application to German health care data on the treatment of very low birth weight infants\cite{gba2021mime}. The latter analysis was done by the Federal Institute for Quality Assurance and Transparency in Healthcare (IQTIG), the central institution for statutory quality assurance in the German public healthcare system. The IQTIG conducts volume-outcome analyses and threshold-value analyses in order to provide statistical advice for the Federal Joint Committee (Gemeinsamer Bundesausschuss, G-BA), which is responsible for stipulating the German minimum caseload requirements. Such analyses motivated the development of the methodology presented in this paper.

The paper is organized as follows. Section 2 introduces notation, visualizes our assumptions on the relationship between the relevant variables in the data generating process in a causal DAG and describes our statistical approach in detail. 
Sections 3 and 4 are devoted to illustrating our approach by way of a simulation study and the application to German health care data on the treatment of very low birth weight infants.
The paper closes with a discussion on the aspect of interventions and gives an outlook.

\section{Flexible Volume-Outcome Analyses} 
\label{sec:voloutana}

Our approach relies on the availability of data for individual patients as opposed to the case where researchers can only access and analyse aggregated data (on the level of providers); see Section \ref{ssec:aggregated} for a discussion.
Based on the individual patient data, we model the probability for the individual binary outcomes.

Let $I$ denote the number of providers under consideration, and let $n_i$ be the number of patients of provider $i\in\{1,2,\dots,I\}$ within the time period under consideration.
We denote the unknown probability of the considered binary outcome for patient $j\in\{1,2,\ldots,n_i\}$ treated by provider $i\in\{1,2,\ldots,I\}$ as $\pi_{ij}$. The corresponding outcome is given by $Y_{ij}\sim\mbox{Ber}(\pi_{ij})$ and the total number of available observations by
$N = \sum_{i=1}^I n_i$.  
Our main interest consists in detecting the relation between the provider volume $\vol_i$ and the probability $\pi_{ij}$.
We first need to specify precisely what we mean by volume.
In the simplest case, one may use the caseload $n_i$ as the volume.
As mentioned in Section~\ref{sec:intro}, the volume can be interpreted as a proxy for underlying unobserved properties of the provider. While the observed caseload inevitably fluctuates between years, one may assume that the underlying unobserved properties (such as staff experience or equipment) do not fluctuate to the same extent. Therefore one may average the caseload over several years in order to obtain a more stable proxy.
We discuss this in detail in the example in Section~\ref{sec:real_world_example}. 
Another definition of volume that is sometimes used is the provider bed count\cite{george2017}.

\subsection{Visualisation in a causal DAG}\label{ssec:dag}~\\
As a first step towards a suitable statistical model, we draw a causal directed acyclic graph (DAG)\cite{pearl2009causality} that visualizes relevant variables in the data generating process in terms of the research question and assumptions on their causal dependencies. Although regression analyses have a clear direction in the sense that one variable is regressed on other variables, they alone---without further assumptions---purely allow statements on associations between variables. However, associations are not necessarily helpful for decision makers when it comes to interventions as the different mechanisms that may lead to volume-outcome associations demonstrate (see Section \ref{sec:intro}). Instead, the interest commonly lies in the causal effect of volume of the past on future outcomes as it is particularly relevant for decision makers whether one can improve the outcomes by intervening on the volume.
Albeit a simplified representation, the DAG presented here helps us to make our assumptions transparent and to decide which variables should be included in the regression model in order to allow interpretations beyond associations. 

\begin{figure}[h]
\centering
\includegraphics[width=0.3\textwidth]{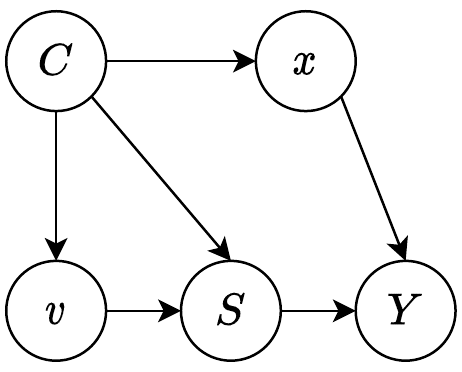}
\caption{DAG for volume-outcome relationship.} \label{fig:DAG}
\end{figure}

Figure \ref{fig:DAG} shows the DAG for the assumed underlying data generating process. To keep the DAG as clear as possible, we only include nodes and edges that are necessary for addressing the question of a volume-outcome relationship. Node $v$ denotes the provider's volume of the past. The provider characteristics are included in node $\boldsymbol{S}$. They consist of both volume-associated and non-volume-associated characteristics related to the provider's treatment quality, e.g. practical experience, equipment and previous performance. 
Node $\boldsymbol{x}$ includes the patient's characteristics (called risk factors in the following), observable or not, such as pre-existing illnesses, that both influence the patient's outcome $Y$ and are related to the choice of a provider. 

The choice is represented by $\boldsymbol{C}$ which contains variables that influence which patient (more precisely with which risk factors $\boldsymbol{x}$) is treated by which provider (with which volume of the past $v$ and characteristics $\boldsymbol{S}$).
Node $\boldsymbol{C}$ thus represents a common cause of $v$, $\boldsymbol{x}$ and $\boldsymbol{S}$ that explains correlation between them. For example, one often observes that smaller providers mainly see patients with lower risks, while high risk patients tend to be treated in larger hospitals\cite{birkmeyer2002hospital,LeeLin2007:volout_psych,george2017,krautz2018effect}.
Thus, $\boldsymbol{C}$ contains information on the region with its demographics and provider landscape and also on medical guidelines that control how high-risk patients are deferred to better-equipped providers.

The edge from volume $v$ to $\boldsymbol{S}$ includes, e.g., the effect of practical experience on the provider's treatment quality. The patient's outcome $Y$ is influenced by both the patient-specific risk factors $\boldsymbol{x}$ as well as the provider's characteristics $\boldsymbol{S}$. 

Based on the interest in the causal effect of volume on outcome, we see that several paths lead from the provider's volume of the past, $v$, to the patient's outcome, $Y$.
Only the directed paths (in which all edges point forward) describe a causal effect.
There is a causal path $v\to\boldsymbol{S}\to Y$ that covers the explanation of \textit{practice-makes-perfect} or can be explained by, e.g., more \textit{specialized equipment} of providers with higher volume leading to better treatment results. Note that this path describes only indirect causal effects from volume $v$ via $\boldsymbol{S}$ on $Y$, and the volume itself is only a proxy, which gains importance for decision makers when the provider characteristics $\boldsymbol{S}$ are not directly accessible for interventions (see Section~\ref{sec:intro} and the discussion on consequences for interventions in Section~\ref{sec:discussion}).

All non-causal paths between $v$ and $Y$ go through the common cause $\boldsymbol{C}$. The paths via both $\boldsymbol{C}$ and $\boldsymbol{S}$ cover---among other things---the \textit{selective-referral} explanation according to which the choice of a provider depends on its characteristics (including former outcomes) leading to higher volumes for providers with better performance.

In order to unbiasedly estimate the causal effect of $v$ on $Y$, we would thus aim to account for all common causes in $\boldsymbol{C}$, which is difficult or even impossible in practice.
We can, however, try to account for the relevant risk factors $\boldsymbol{x}$ that are both related to the choice where the patient is treated and have an effect on the patient's outcome.

Suppose that we can account for these patient's risk factors. Then the only remaining non-causal path from $v$ to $Y$ that may induce correlation goes through first $\boldsymbol{C}$ and then $\boldsymbol{S}$. 
This path can only be neglected if one can assume that the observed correlation between $v$ and $Y$ can largely be traced back to the causal effect of $v$ on $\boldsymbol{S}$ (e.g. more practice, change of internal processes, need and money to purchase specialized equipment) rather than to the common cause $\boldsymbol{C}$ (e.g. changes in provider landscape or selective referral). If this is the case the arrow from $\boldsymbol{C}$ to $\boldsymbol{S}$ could be omitted.

To sum up, whether the estimated effect of volume on outcome can be interpreted causally thus depends on the one hand on whether all relevant patient risk factors are taken into account and on the other hand on the relevance of the non-causal association between $v$ and $\boldsymbol{S}$ compared to the causal effect of $v$ on $\boldsymbol{S}$.

\subsection{Statistical Model}\label{sec:model}~\\
As a main benefit from constructing the DAG, we know which factors potentially cause confounding of the volume-effect and should therefore preferably be incorporated in the statistical analysis. Further decisions in setting up the concrete regression model concern the outcome distribution and specific modelling of the individual effects.  
\bigskip

As we focus on a binary outcome, a logistic model for the probability $\pi_{ij}$ suggests itself. In particular, consider
\begin{equation}\label{eq:abstract_model}
\logit(\pi_{ij}) = \eta_{ij} + b_i,
\end{equation}
where the effect $\eta_{ij}\in\mathbb{R}$ depends on the observed patient-specific risk factors $\widetilde{\boldsymbol{x}}_{ij}$ (including the global intercept) and $b_i\in\mathbb{R}$ on the provider-specific factors in $\boldsymbol{S}$.
Note the difference between the vector $\boldsymbol{x}$ of all risk factors that are related to both the choice of a provider and the patient's outcome $Y$ in the DAG and the vector $\widetilde{\boldsymbol{x}}_{ij}$ of those that are actually observed and can be included in the model.

When selecting the patient-specific risk factors and when modelling the functional form of $\eta_{ij}$, the general rules of confounder control and risk adjustment apply regarding the choice and modelling of risk factors\cite{Iezzoni2013:RA_for_Healthcare,Fahrmeiretal2021:Regression}. 

In the simplest case, $\eta_{ij}$ may be a linear predictor of the form
\begin{equation}
\label{eq:etaij}
\eta_{ij} = \widetilde{\boldsymbol{x}}_{ij}^\top \boldsymbol{\beta}.
\end{equation}
In more complicated situations, $\eta_{ij}$ may include smooth terms for continuous patient-specific risk factors or interactions between different risk factors.

With regard to the provider-specific component in \eqref{eq:abstract_model}, we propose the decomposition 
\begin{equation}\label{eq:bi}
b_i = \fvol (\vol_i) + u_i,
\end{equation}
where $\fvol$ is the smooth volume effect modelled through penalized splines and $u_i\in\mathbb{R}$ is a volume-independent provider-specific effect modelled as a random intercept. These two effects are discussed in detail in Sections~\ref{ssec:volumeeffect} and~\ref{ssec:randomintercept}. Note that, in light of the discussion of Section \ref{ssec:dag}, $\fvol$ incorporates the causal effect of $v$ onto $\boldsymbol{S}$ as well the potential non-causal effect of $v$ onto $\boldsymbol{S}$ via $\boldsymbol{C}$. 

One may also include further provider-specific covariates, if such information is available---for instance, a parameter for teaching hospitals or parameters for certain structural properties such as the availability of certain equipment or staff.
Note that including further provider-specific covariates can change the volume effect $\fvol$ as well as the volume-independent provider-specific effect $u_i$. 

In sum, our model becomes
\begin{equation}\label{eq:standard_case}
\logit(\pi_{ij}) = \eta_{ij} + \underbrace{\fvol (\vol_i) + u_i}_{=b_i}.
\end{equation}

The observations are assumed to be conditionally independent given the included covariates:
$$Y_{ij}~|~\widetilde{\boldsymbol{x}}_{ij}, \vol_i, u_i~\ind~\mbox{Ber}(\pi_{ij}).$$
The model's conditional nature also determines how to interpret the results. For instance, one has to be aware that adding covariates for provider properties affects the interpretation of $\fvol$. With additional covariates, $\fvol$ describes how the risk depends on the volume for providers coinciding with respect to those covariates. On top of that, again, $\fvol$ can only be interpreted as the causal volume effect if $\widetilde{\boldsymbol{x}}_{ij}$ and $\boldsymbol{x}_{ij}$ do not differ essentially (i.e., if all relevant risk factors are observed) and if the relevance of the non-causal association between $v$ and $\mathbf{S}$ is negligable compared to the causal effect of $v$ on $\mathbf{S}$.

\bigskip

Depending on the specific application, Model~\eqref{eq:standard_case} can be extended in different ways without affecting our methodology. For instance, when data about longer time periods are available, one may have to decide whether any of the parameters has to be modelled as time-dependent. We present an example in Section~\ref{sec:real_world_example}.

Aside from that, the model may be extended by introducing interactions between the provider-specific factors $u_i$ or $\fvol (\vol_i)$ and the patient-specific risk factors. The assumption that such interactions are not relevant is often made, mostly because of technical convenience and because it is difficult to estimate a large number of interaction parameters. It has been found that the assumption still leads to a good model, as long as interactions are not too strong \cite{VarewyckVansteelandtErikssonGoetghebeur16:center-patient-interactions}. However, if it turns out that interactions between the volume $\vol_i$ and the patient-specific risk factors $\widetilde{\boldsymbol{x}}_{ij}$ are relevant, then the volume effect may depend in a strong way on the patient-specific risk factors; that is, a uniform volume effect for the outcome and patient population under consideration does not exist. 
In that case, not all patients profit equally from a potential volume-dependent intervention (and some patient groups may even be harmed if the direction of the effect depends on patient characteristics).

\subsection{The volume effect}\label{ssec:volumeeffect}
	
A central aspect of our approach consists in modelling the volume effect $\fvol$ as a smooth function in $\vol_i$.
This makes it possible to detect a wide range of possible effects, including monotonic and non-monotonic ones.
The functions' smoothness reflects the assumption that the effect of volume on the outcome probability would not abruptly change once the volume takes a slightly higher or lower value. We penalize the wiggliness of the estimated function $\widehat{\fvol}$ by a penalty term controlled by a smoothness parameter in order to cope with overfitting\cite{eilers1996flexible}. The smoothness parameter is jointly estimated along with all other model parameters as part of the unified estimation procedure\cite{wood2017}.

In contrast to that, the common practice of using volume groups restricts the volume effect to be a step function based on a rather arbitrary partition.  This dramatically limits the range of possible shapes and inevitably causes unrealistic jump discontinuities. Furthermore, results may exhibit a strong sensitivity regarding the chosen partition, i.e. the specific boundaries between volume groups\cite{GrouvenKuechenhoffSchraederBender08:Review_minimum_provider_volumes}, which can also be seen in examples\cite{birkmeyer2002hospital, mikeljevic2003surgeon, varagunam2015relationship}.

To overcome the limitations of discrete volume groups, Nimptsch and Mansky\cite{NimptschMansky17:25Volout} redo their analyses with a linear volume effect. They observe that in some cases, a volume effect that appears significant when the volume is discretized is no longer significant when volume is treated linearly.
They fail to conclude that their results provide evidence that a linear model is inappropriate in these examples.
This example demonstrates that a more flexible model is needed in general\cite{christian2005}.
The penalized spline approach also requires certain choices, albeit much less consequential ones\cite{eilers1996flexible}. 

When specifying the spline, the particular distribution of observed provider volumes deserves special attention. 
In the context of splines, different approaches exist to take into account imbalances in this distribution\cite{wood2003, wood2017}: bases with quantile-based knots, knot-free splines or adaptive smoothing. In extreme cases it may be necessary to transform the provider volumes (e.g. $\log$-transformation) \cite{george2017}. Employing such strategies is recommendable in order to cope with uncertainty in regions with little or no data on the one hand and prevent oversimplification in regions with many observations on the other hand. 

Of course, other continuous patient-specific risk factors (or provider properties) should be treated analogously.

\subsection{The provider-specific random intercept} 
\label{ssec:randomintercept}

The provider-specific random intercepts explicitly model dependence between observations from a common provider, i.e. clustering. They directly represent differences between providers that are not covered by the volume effect or other available provider properties.
However, they have to be interpreted carefully, because they may under certain circumstances also depend on patient properties: If there are relevant unobserved risk factors 
that are unevenly distributed among the providers, their effect is also picked up by the random intercepts\cite{CMS_whitepaper2012}.
When clustering of providers is ignored, the importance of the volume effect tends to be overestimated\cite{urbach2005}.

We assume that the random intercepts are normally distributed,
$$u_i\iid\mathcal{N}(0,\tau^2),$$
and stochastically independent from all other model covariates. 
These typical assumptions\cite{stroup2012, gelman2006} are generally not too restrictive \cite{mcculloch2011, townsend2013}. 

The associated standard deviation $\tau>0$ is estimated along with the other parameters and deserves some attention: It explicitly quantifies the heterogeneity between providers beyond the heterogeneity that can be explained by variations in volume and other provider-specific as well as patient-specific effects incorporated in the model.
This enables direct comparisons between the two components of the provider-specific effect $b_i$ (see~\eqref{eq:bi}),
i.e. the volume effect $\fvol$ and the volume-independent effect represented by the $u_i$.

In principle, other methods for taking into account clustering exist. Instead of random intercepts one may consider fixed effects.
However, fixed effects may cause identifiability problems as $\fvol$ also represents a fixed effect on the provider level\cite{townsend2013}, and they may be numerically less stable when there are small providers without adverse events. One solution to this is to apply Firth regression to estimate penalized provider-specific fixed effects\cite{firth1993bias,varewyck2014shrinkage}. 
Another alternative to conditional modelling through random intercepts consists in marginal modelling where the dependence of outcomes of patients treated by the same provider is accounted for in the covariance structure rather than in the mean \cite{fitzmaurice2011,NimptschMansky17:25Volout,iqwig06:MiMe_KCH,panageas2003}.

However, the random intercept approach exhibits an important advantage over both these alternatives: It directly yields an estimate of the valuable parameter~$\tau$ which has the natural interpretation of measuring the between-provider heterogeneity. 
Furthermore, conditional modelling (as opposed to marginal modelling) better suits the objective of measuring a volume effect that is adjusted for between-provider and between-patient differences that are independent of differences in volume.

The question remains how to specify the clusters. 
For instance, one could group patients with respect to the attending physician rather than the hospital.
Such a choice very much concerns the scientific question since individual physician volume and hospital volume represent different aspects\cite{christian2005}. While physician volume strongly emphasizes individual experience and human factors, the availability of technical equipment may correlate more with hospital volume, for instance. 
However, even if the more detailed information is available and of interest, this may lead to a very high number of random intercepts $u_i$ with potentially imprecise estimates.

\subsection{Statistical Inference}\label{sec:stat_inf}

As part of the volume-outcome analysis, a statistical test for the hypotheses
\begin{equation}H_0:~\fvol \equiv \mbox{const.}~~\mbox{vs.}~~H_0:~\fvol \nequiv \mbox{const.}\label{eq:test_problem}\end{equation}
can be performed which corresponds to testing if the volume effect is zero/non-existent, against the alternative that there is an effect somewhere between the lowest and highest observed volumes. 
A suitable test statistic as well as associated confidence bands based on a Bayesian perspective of the model have already been derived for GAMMs; for more details we refer to the relevant literature\cite{marra2012, wood2012}. Furthermore, a test on the existence of the random intercepts, i.e. a test for the null hypothesis $\tau = 0$ versus the alternative $\tau > 0$, is available\cite{wood13}.

We propose to compare the volume effect $\fvol$ and the volume-independent provider effect $u_i$ through evaluating odds ratios with respect to the volume on the one hand and the median odds ratio of the random intercepts on the other hand\cite{larsen2000interpreting}.

We first elaborate on the volume effect. Consider two (hypothetical) providers $i_1,i_2$ with the same random intercept, $u = u_{i_1} = u_{i_2}$, but different volumes $\vol_1 \neq \vol_2$. Consider two patients, one that is treated by provider $i_1$ and the other by $i_2$, and assume that the patients are identical with respect to all patient-specific covariates $\widetilde{\boldsymbol{x}}$ included in the model. Then, the odds ratio comparing the odds of the outcome for the patient treated by provider $i_1$ with those of the patient treated by provider $i_2$ is given by
\begin{align}
\mbox{OR}(\vol_1,\vol_2) &= \frac{\mbox{Odds}(Y = 1~|~\widetilde{\boldsymbol{x}}, \vol_1, u)}{\mbox{Odds}(Y = 1~|~\widetilde{\boldsymbol{x}}, \vol_2, u)}\nonumber \\ &= \exp\big(\fvol(\vol_1)-\fvol(\vol_2)\big).\label{eq:OR_for_volume}
\end{align} 
Note that $\mbox{OR}(\vol_1,\vol_2)$ compares the volume effects adjusted for provider characteristics ($u$) and patient characteristics ($\widetilde{\boldsymbol{x}}$).
Estimates $\widehat{\mbox{OR}}(v_1,v_2)$ can be obtained as plug-in estimates. In the supplemental material, Section \ref{sec:or_est}, we provide an algorithm to obtain confidence intervals for $\mbox{OR}$.

Now, given odds ratios for varying volumes, we require a quantity describing the volume-independent provider effect allowing for a direct comparison. The \textit{median odds ratio} (MOR)\cite{larsen2000interpreting} based on the standard deviation of the random intercepts, $\tau$, provides just that. 
It is defined as
$$
\mbox{MOR} = \exp\left(-\sqrt{2}\Phi^{-1}(\sfrac{3}{4})\tau\right),~~\tau>0.
$$
The MOR always lies in $(0,1)$ and can be interpreted as follows: Consider two patients which are identical with respect to their risk factors in the model. Assume that the patients are randomly assigned to providers with the same volume but (potentially) different random intercepts. This yields a random odds ratio for the outcome of interest between the patient treated by the lower risk provider (smaller random intercept) and the patient treated by the higher risk provider. 
The median of this random odds ratio's distribution is the $\mbox{MOR}$. Hence, the $\mbox{MOR}$, just as $\tau$, quantifies the magnitude of the volume-independent provider effect as measured through the random intercepts without depending on other properties of the patient or provider.
A plugin-in-estimate can be obtained based on the estimate $\widehat{\tau}$. 
Due to the strict monotonicity of the $\mbox{MOR}$ as a function of $\tau$, confidence intervals for $\tau$ directly yield confidence intervals for the $\mbox{MOR}$ by transforming the limits. 

\subsection{Using aggregated data}\label{ssec:aggregated}

Whenever one lacks data on the individual patient level, the analysis will have to rely on aggregated results on the provider level\cite{rogowski2004indirect,Heller2018:Regionalisierung}. At best, those results are already adjusted for patient-specific risk factors. For instance, that would be the case for standardized mortality ratios (SMRs)\cite{hosmer2013}. Of course, such an approach requires no consideration of clusters. 
 
Clearly, from a purely statistical point of view, working with aggregated results is suboptimal and comes along with a loss of information\cite{christian2005}.
For instance, with aggregated results it is usually not possible to properly adjust for confounding by patient-specific risk factors since patient-specific effects and the volume effect are estimated in two separate steps rather than simultaneously in a joint model. 
Also, it is difficult to incorporate the uncertainty of the risk adjustment procedure into the analysis.  
In any case, one has to take into account that the aggregated results of smaller providers have more statistical variability than those of larger providers, which may be addressed by down-weighting the data from smaller providers\footnote{In comparison, when working with individual patient data, each data point carries the same weight, but clustering of the data into providers must be taken into account.}. 
That being said, aggregated results exhibit the the advantage that they are often more easily available.

\subsection{Software}
\label{sec:software}

Algorithms to fit GAMMs are implemented in all major statistical software packages.  For $\mathsf{R}$, we recommend the flexible, fast and well-validated package \texttt{mgcv} \cite{R-project,wood2017,wood2021}. It offers a penalized maximum likelihood approach very well suited to estimate all quantities (with confidence sets) described in this section in a unified fashion; including the smoothness parameter.
The source code for the simulation study in Section~\ref{sec:simulations} is available on GitHub\cite{github_repo} and demonstrates how our approach can be implemented using \texttt{mgcv}.

In principle, many choices of spline bases and penalty approaches exist\cite{perperoglou2019review}. We recommend to study their impact on the overall results in sensitivity analyses, especially if the provider volumes $v_i$ are unevenly spread.

\section{Simulations}
\label{sec:simulations}

We apply our estimation approach in a simulation study.
The goal of the simulation study is to apply our methods to a setting with parameter values that we consider realistic and to observe how reliably our methods detect the volume effect as the number of hospitals and the underlying shape of the volume effect varies.
We include two patient-specific risk factors: one binary factor and one continuous factor with a linear effect (yielding a simple linear predictor $\eta_{ij} = \widetilde{\boldsymbol{x}}_{ij}^\top\boldsymbol{\beta}$, see \eqref{eq:etaij}).
The parameters of the simulation are summarized in Table~\ref{tab:simparams}.
The $\mathsf{R}$-code for the simulation and its analysis is available at GitHub\cite{github_repo}.
\begin{table}[ht]
\begin{tabular}{ccp{0.5\textwidth}}
\toprule
parameter & value & description \\
\midrule
$I$       & (see text) & the number of providers \\
$\mu_n$   & 100 & the average volume of the providers \\
$\fvol$   & (see text) & the volume effect \\
$\tau$    & 0.5 & the standard deviation of the provider effects \\
$\pi_1$   & 0.3 & the prevalence of the binary risk factor \\
$\beta_0$ & $\operatorname{logit}^{-1}(0.1)$ & intercept \\
$\beta_1$ & 0.3 & coefficient of the binary risk factor \\
$\beta_2$ & 0.5 & coefficient of the linear risk factor \\
\bottomrule
\end{tabular}
\caption{The parameters of the simulation.}
\label{tab:simparams}
\end{table}

Given the parameters, we generate data according to the following algorithm.
For each provider $i=1,\dots,I$:
  \begin{enumerate}
  \item Draw the caseload $n_i\sim\text{Poisson}(\mu_n)$.
  \item Draw the provider effect $u_i\sim \mathcal{N}(0,\tau^2)$.
  \item For each patient $j=1,\dots,n_i$:
    \begin{enumerate}
    \item Draw the discrete risk factor $x_{ij1}\sim\text{Bernoulli}(\pi_1)$.
    \item Draw the continuous risk factor $x_{ij2}\sim \mathcal{N}(0,1)$.
    \item Draw the outcome according to
    \end{enumerate}
\begin{equation*}
\mathds{P}(Y_{ij} = 1~|~u_i, n_i, x_{ij1}, x_{ij2}) %\\
= \operatorname{logit}(\beta_0 + \beta_1 x_{ij1} + \beta_2 x_{ij2} + \fvol(n_i) + u_i).
\end{equation*}
  \end{enumerate}
After simulating the data, we fit Model \eqref{eq:standard_case}, where we use the caseload as the volume ($\vol_i=n_i$).

In this section, we focus on the case of a U-shaped volume-outcome relationship: $\fvol(n) = 1/1000 (n - 100)^2$. The supplemental material (Section~\ref{sec:more_simulations}) contains results for the case of no volume-outcome relationship and for the case of a linear volume-outcome relationship.
The results in the linear case are similar to the results in the U-shaped case.
The case of no volume-outcome relationship confirms that the false discovery rate can be controlled by the significance level in the sense that, for instance, the $p$-value of the test for a volume effect lies below 0.05 in approximately 5\% of the simulation runs. 
Moreover, in this section, we keep the standard deviation of the provider effects $\tau$ fixed at $0.5$.
Results for other values of $\tau$ are presented in the supplemental material (Section~\ref{sec:more_simulations}).  Those results are quite similar; however, when $\tau$ becomes larger, it becomes more difficult to reliably detect a volume effect.

\begin{figure}
\centering
a)\raisebox{-0.4\textwidth}{\includegraphics[width=0.45\textwidth,height=0.45\textwidth]{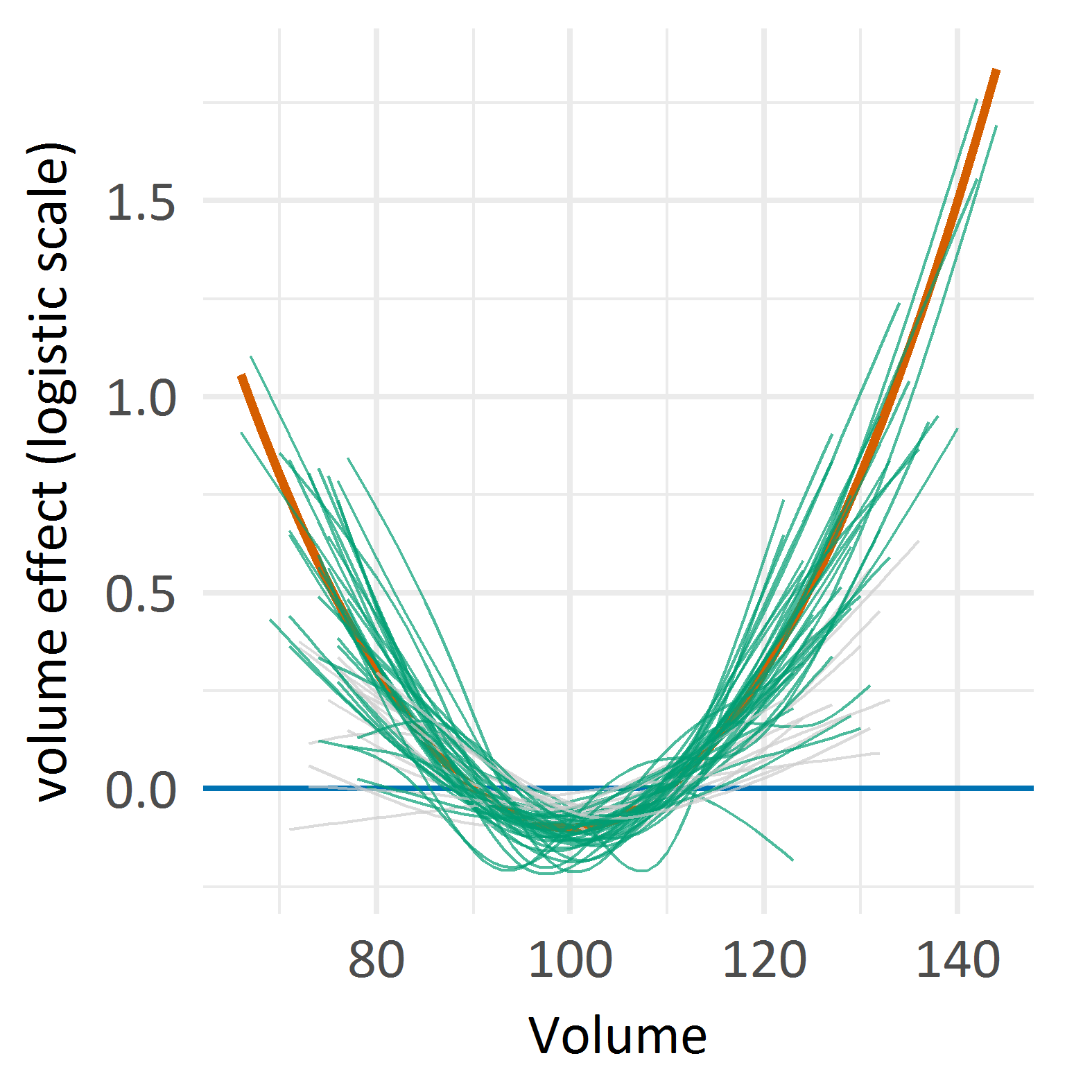}} 
b)\raisebox{-0.4\textwidth}{\includegraphics[width=0.45\textwidth,height=0.45\textwidth]{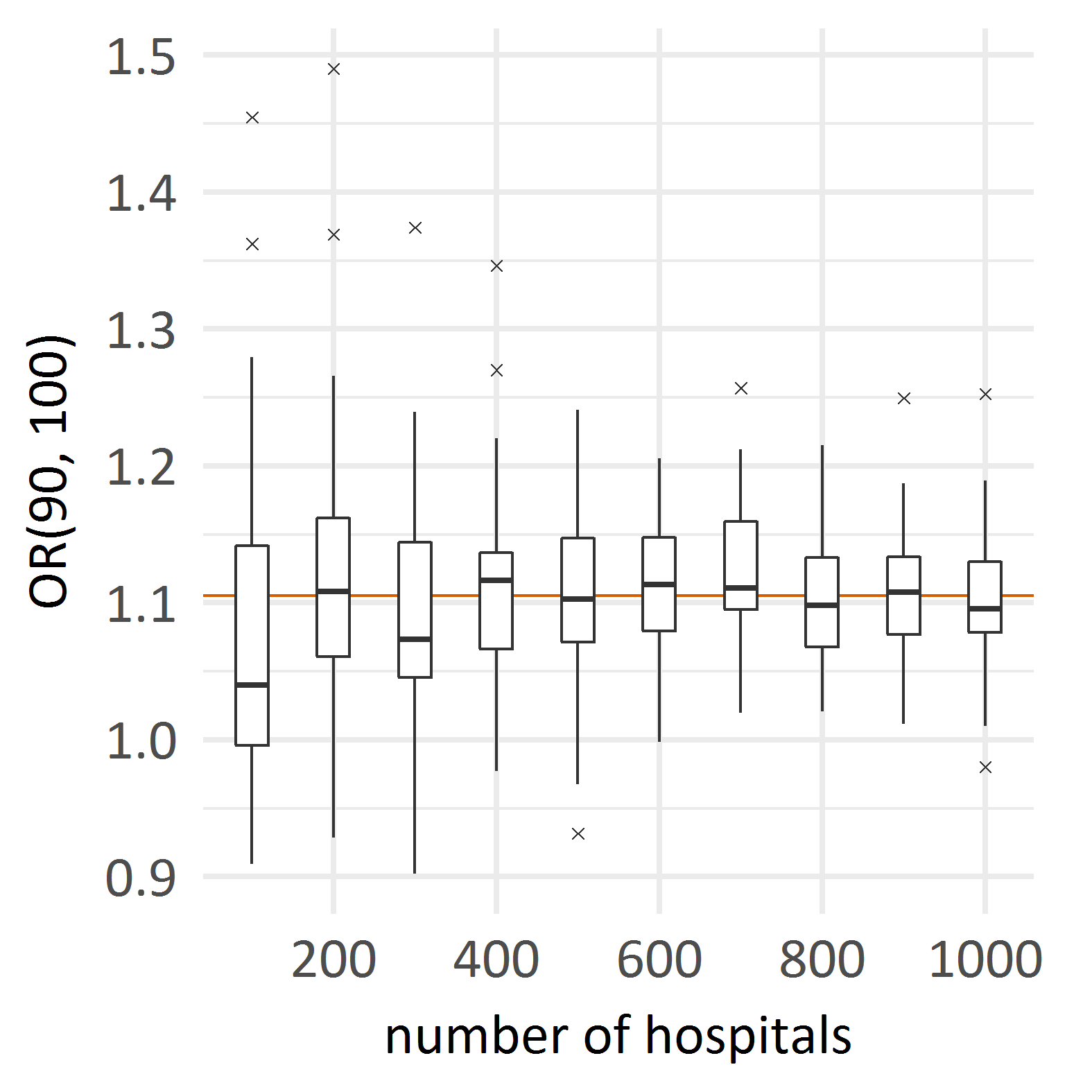}}
\caption{a)~Illustration of the estimated volume effect $\widehat{\fvol}$ in 50 simulation runs with a true U-shaped volume-outcome relationship (in red), $I=200$ and $\tau=0.5$. Curves are green if the $p$-value of the test for a volume effect is $\le 0.05$.\label{fig:simpredplotall}
b)~Estimates of the odds ratio OR(90, 100) obtained in several simulation runs. For each number of hospitals, 50 runs were made.
The true value is plotted in red.\label{fig:simORplot}}
\end{figure}

Figure~\ref{fig:simpredplotall}a) shows the estimated volume effect for 50 simulation runs with $I = 200$ providers. The estimated volume effects lie reasonably close to the true volume effect (red line).
Figure~\ref{fig:simpredplot} in the supplemental material shows the estimate of the volume effect for four of these simulation runs, together with confidence bands.
To illustrate the estimation of odds ratios ${\mbox{OR}}(v_1,v_2)$, we focus on the exemplary odds ratio OR(90,100).
Figure~\ref{fig:simORplot}b) shows the distribution of the estimates of the odds ratio OR(90,100) for 500 simulation runs for $I=100, 200, \dots, 1000$.
The variation of the estimates around the true value (horizontal red line) is reasonably small, at least when the number of hospitals is not too small.
Figure~\ref{fig:ORplot_sel_v2} in the supplemental material shows estimates with confidence intervals of $\mbox{OR}(90,100)$ for 200 simulation runs and illustrates how confidence intervals become smaller as $I$ increases. 

\begin{figure}
\centering
a)\raisebox{-0.4\textwidth}{\includegraphics[width=0.45\textwidth,height=0.45\textwidth]{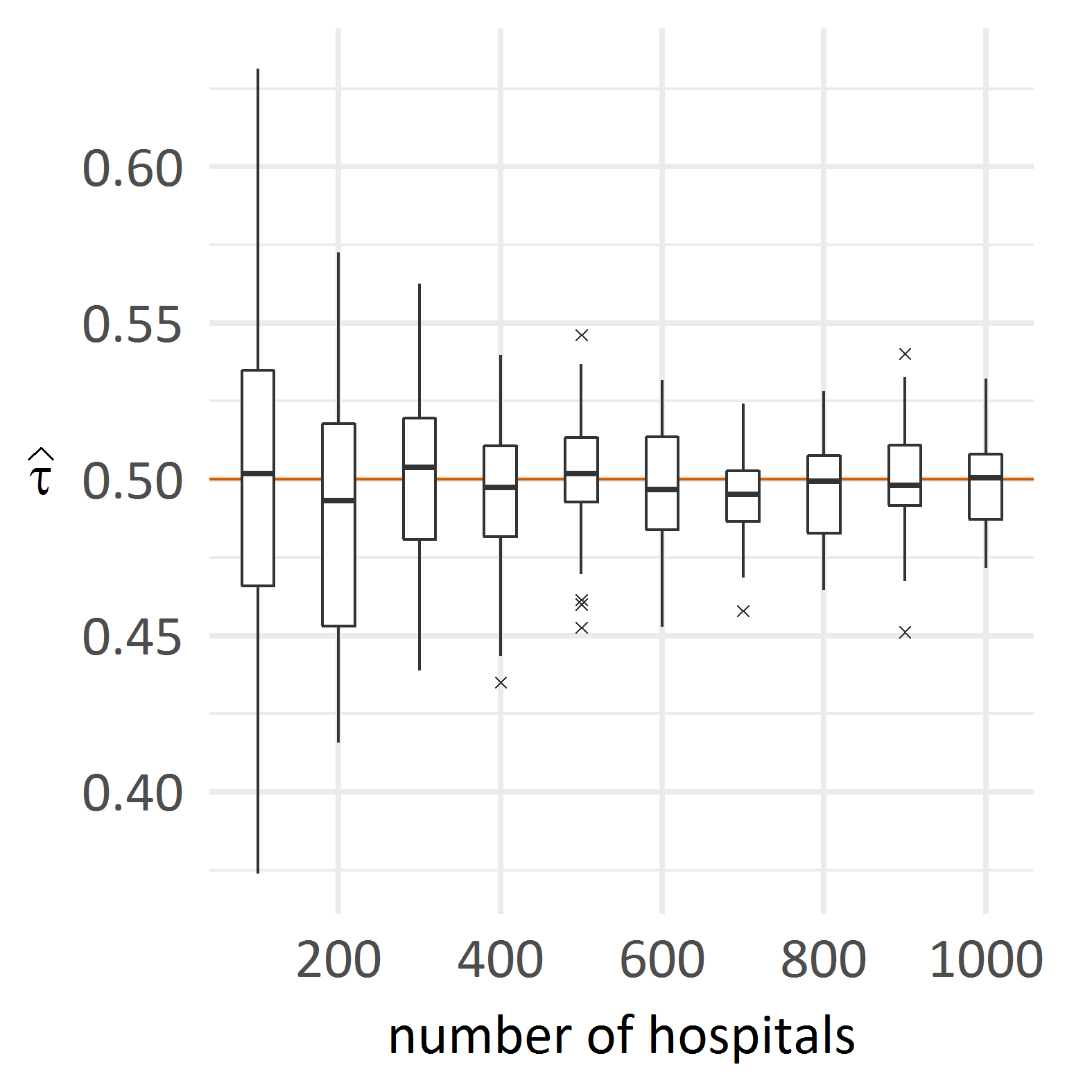}}
b)\raisebox{-0.4\textwidth}{\includegraphics[width=0.45\textwidth,height=0.45\textwidth]{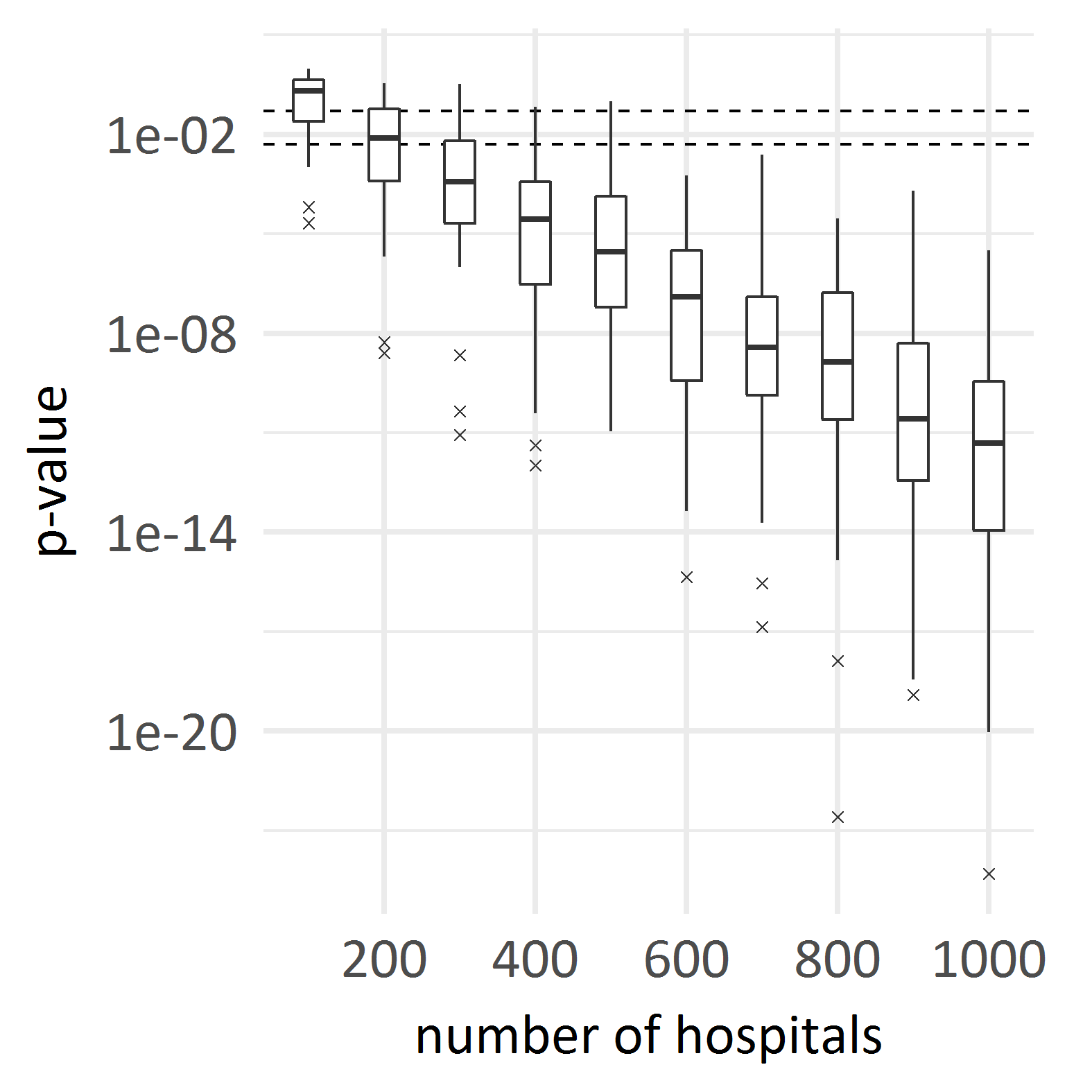}}
\caption{
  Results from 500 simulation runs (50 runs for each number of hospitals).
  a)~The estimate $\widehat\tau$. The horizontal line marks the true value. \label{fig:simtau}
  b)~The $p$-value of the test for a volume effect. The dashed horizontal lines mark the canonical significance levels 0.05 and 0.005. \label{fig:simps}}
\end{figure}

Figure~\ref{fig:simtau}a) shows the estimated value of~$\tau$. The estimates $\widehat\tau$ lie close to the true value when $I$ is large enough.
Figure~\ref{fig:simps}b) shows the $p$-value of the test for a volume effect.
The $p$-values spread more when the number of hospitals is small.  This suggests that, in a setting that closely matches the parameters of our simulation, it becomes difficult to reliably detect the volume-outcome relationship when the number of hospitals is 200 or smaller.
Of course, detecting a volume-outcome relationship may still be possible with fewer hospitals if either the effect is stronger or if data from several years are available (as in the example in Section~\ref{sec:real_world_example}). 
In contrast, the provider effect was always significant ($p$-value less than $10^{-9}$ in all simulation runs).

\section{Application to hospital data on very low birth weight infants}\label{sec:real_world_example}

\subsubsection*{Background.}
The retrospective analysis aims at estimating the volume effect $\fvol$ in the treatment of very low birth weight (VLBW) infants, i.e. newborns with birth weights below 1250g. 
The outcome of interest is death and the providers correspond to the German neonatal intensive care units (NICU). 

\subsubsection*{Data.}
The study uses data from the QFR program\cite{QFR_Richtlinie_2005}, which is maintained by the IQTIG and, among other things, runs a public website that summarizes treatment outcomes of VLBW infants at German NICUs\cite{perinatalzentren.org}. Our study is based on the data from the years 2014--2018, the available database consists of 29,048 documented infants.
This essentially covers all VLBW infants born in Germany during these years.
When fitting the model, we exclude palliative care patients and infants born before the 24th week of gestation as these patients are not considered suited for measuring quality of care. After this exclusion, 26,043 patients remain in 163 providers. 
The resulting population exhibits an overall death rate of about 7.3\%.
However,we use the full database of 29,048 patients when computing the providers' volumes~$\vol_i$, since these patients also count towards the minimum caseload requirements\cite{MiMeRegelungen2021} and since one may presume that these patients also add to the experience of the staff.

\subsubsection*{Model.}
The provider's volume can be operationalized in different ways: Defining $\vol_i$ as a simple five-year-average volume is an easy approach; alternatively, one may associate each patient with the volume of the respective provider within the respective year, yielding a time-dependent version of $\vol_i$.

In the case of the VLBW-study, we use a construction that is primarily guided by two aspects:
Firstly, from a causal perspective, future patients cannot possibly affect a present patient so that a simple five-year-average appears inappropriate.  For example, the treatment outcome of an infant born in 2014 cannot causally depend on the provider volume in year 2018.
Secondly, as already mentioned in Section \ref{sec:intro}, the volume can be interpreted as a proxy for other properties of the provider.  While the volume inevitably fluctuates between years, one may assume that the underlying properties (such as staff experience) do not fluctuate to the same extent.  Therefore, one may assume that averaging the volume over several years gives a better and more stable proxy. 
Figure \ref{fig:NICU_volume_plot} in the supplemental material depicts the volume fluctuations observed in the VLBW data.

Taking into account both aspects, we use a cumulative average: Say, the analysis is based on observations from consecutive years $0,1,\ldots,h$ and we have access to the volume per year $\widetilde{v}_i^k$ of provider $i$ for the years $k\in\{-s,\ldots,-1,0,1,\ldots, h\}$, with $s\ge 0$ . 
Here, $s>0$ means that historical provider volumes are available. 
The volumes are then defined as

\begin{equation}
\vol_i^k = \left(\displaystyle\sum_{l = -s}^{k} \widetilde{v}_i^l\right)/\left(\displaystyle\sum_{l = -s}^{k} \mathds{1}_{\{\widetilde{v}_i^l > 0\}}\right),~~k\in\{0,1,\ldots,u\}.
\label{eq:stable_proxy}
\end{equation}

Hence, the volume of provider $i$ in year $k$ is the average of all available non-zero volumes per year of provider $i$ up to year $k$. 
Therefore, we use a model that generalizes our standard model~\eqref{eq:standard_case} such that the volume becomes time-dependent.
In the present study, volumes were available from 2012 onwards ($s = 2$).

In total, we arrive at the following model formula for the probability $\pi_{ij}^k$ that the $j$th infant treated by provider $i$ in year $k$ dies in hospital:
\begin{equation}\label{eq:Mime_formula}
\logit(\pi_{ij}^k) = \beta_0^k + \eta_{ij}^k + \fvol (\vol^k_i) + u_i.
\end{equation}
The risk factors $\widetilde{\boldsymbol{x}}_{ij}^k$ were taken from the risk adjustment model of the QFR mortality indicator\cite{NICU_RA2018}. Two of the risk factors (both related to birth weight in grams) were treated as continuous variables with smooth effects just as the provider volume.
Following that same model, we have added a year-dependent intercept $\beta_0^k$ in order to model possible time trends over the five year period. 
The time dependence was weak, and it was not deemed necessary to allow time-dependence for the effects of the patient characteristics. 

We also assume that the volume-independent effect $u_i$ is constant over the years. This reflects the idea that the associated provider properties remain relatively stable and fluctuate less strongly than the caseload over a period of five years. 

Finally, we assume that the volume effect $\fvol$ is time constant. We relaxed this condition in a sensitivity analysis, in which we did not find a strong time dependence. 
Notable fluctuations in $\fvol$ over the years would have to be analyzed and understood before drawing any conclusions about minimum caseload requirements. 

\subsubsection*{Results.}
\begin{figure}
\centering
\includegraphics[width=25pc,height=15pc]{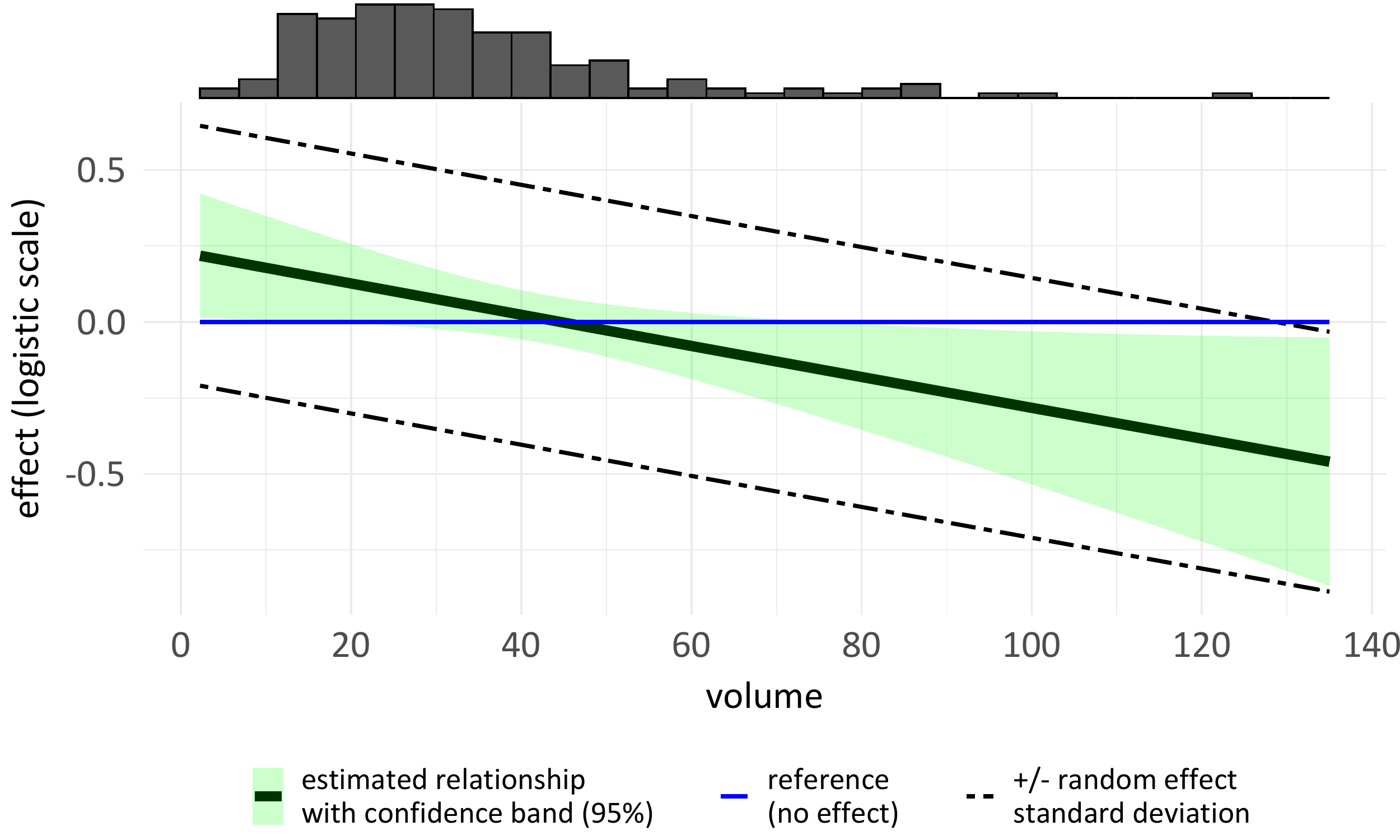}
\caption{Illustration of the estimated volume effect $\widehat{\fvol}$ (solid line) in the VLBW-study on the logistic scale; including a simultaneous 95\%-confidence-band, indication of the volume-independent provider effects' magnitude (dashed lines) and a histogram of the used provider volumes.\label{fig1}}
\end{figure}

Figure \ref{fig1} shows the estimated volume effect $\widehat{\fvol}$, a simultaneous 95\%-confidence-band for $\fvol$ as well as $\widehat{\fvol}\pm \widehat{\tau}$. A sum-to-zero constraint ensures the centering of $\widehat{\fvol}$ around the reference value zero.
The fact that the confidence band does not fully cover the reference line indicates a significant volume effect (level $\alpha = 0.05$) with respect to the testing problem \eqref{eq:test_problem}; the $p$-value in that particular example approximately equals $0.014$. Moreover, the dashed lines suggest a strong influence of the volume-independent provider effects~$u_i$. This influence is quantified through the estimated $\widehat{\mbox{MOR}}\approx 0.665.$
Compared to that, volume-based odds ratios often remain closer to 1 even for relatively high differences $\vol_2-\vol_1$, e.g. 
$$
\mbox{OR}(20,40)\approx 0.902,~\mbox{OR}(20,70)\approx 0.774,~\mbox{OR}(20,100)\approx 0.664.
$$ 
Thus, in this example, volume-independent provider effects appear to exert quite a strong influence on the outcomes compared to the volume effect. The last value marks a constellation of volumes where the volume effect (on the OR scale) approximately equals the MOR, i.e. only a very large increase in volume (from 20 to 100) has an equally strong effect on the outcome as the variability of the volume-independent provider characteristics.

\bigskip

\begin{figure}
\centering
\includegraphics[width=25pc,height=15pc]{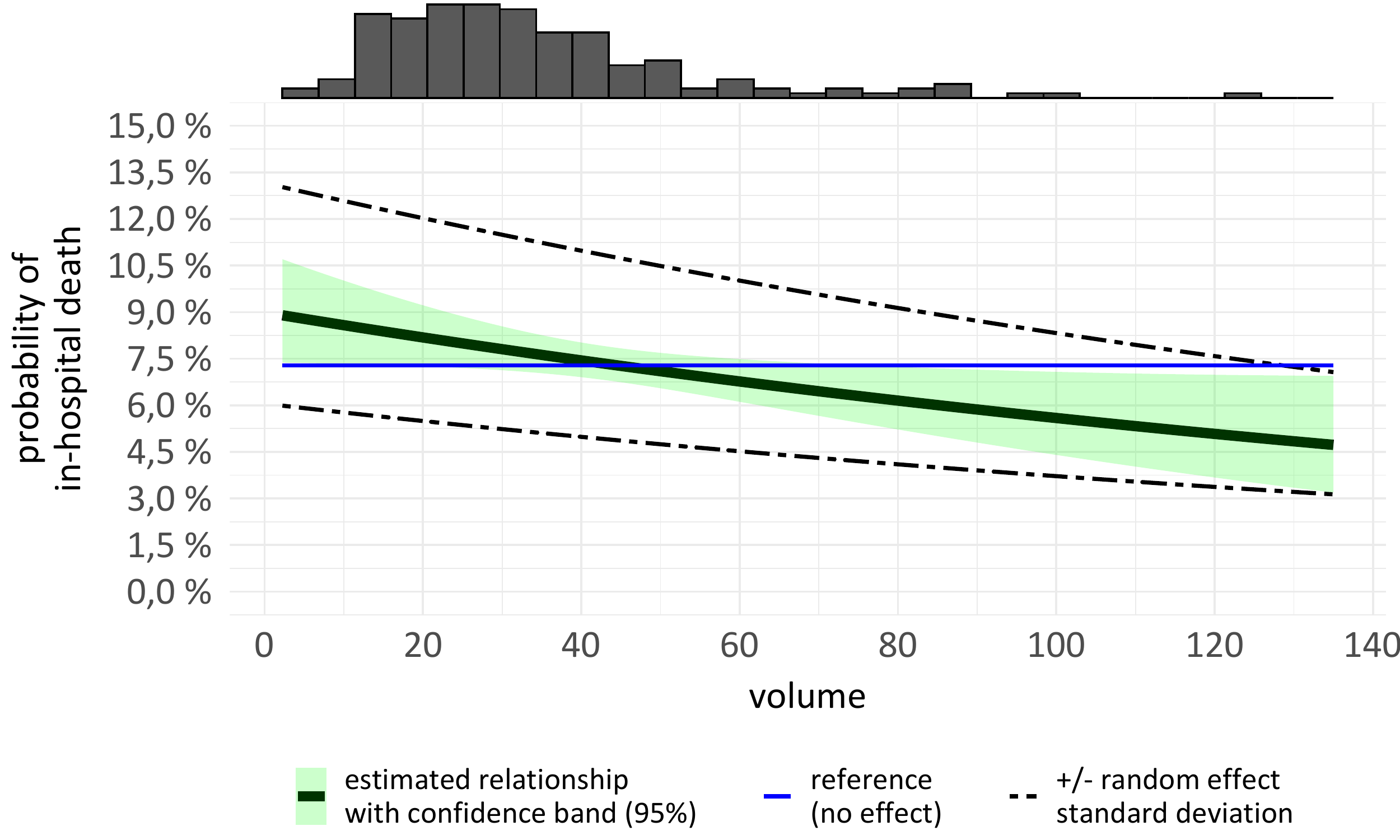}
\caption{Illustration of the estimated volume effect $\widehat{\fvol}$ in the VLBW-study on a probabilistic scale; including a simultaneous 95\%-confidence-band, indication of the volume-independent provider effects' magnitude and a histogram of the used provider volumes.\label{fig2}
}
\end{figure}

Actually, Figure \ref{fig1} does not yet fully address our main objective: understanding the relation between the outcome \textit{probability}, i.e. here probability of death, and the volume effect $\fvol$. In order to obtain a visualisation on the probability scale, we propose the following strategy based on Equation~\eqref{eq:standard_case}. Computing a meaningful probability of the form
$$
\pi(\vol) = \logit^{-1}\left(\eta + \fvol (\vol) + u\right)
$$
requires suitable fixed choices for $\eta$ and~$u$.
If we ensure that these choices represent an average patient, the resulting curve should reside in a range on the probability axis which adequately reflects the observed data. Therefore, let
$$
\widehat{\eta}^\ast = \logit\left(\frac{1}{N}\sum_{i = 1}^{I}\sum_{j = 1}^{n_i} \logit^{-1}\left(\eta_{ij}\right) \right)
$$
and, based on that,
$$
\pi^\ast(\vol) = \logit^{-1}\left(\widehat{\eta}^\ast + \widehat{\fvol}(\vol)\right).
$$
In particular, $u$ is omitted in both expressions as the $u_i$ have zero mean by definition, so that $u = 0$ represents an average provider in terms of its volume-independent effect. The function $\pi^\ast(\vol)$ is depicted in Figure~\ref{fig2}.

Note that such a construction does not add information regarding the statistical evaluation of the volume effect. That is, Figure \ref{fig2} carries essentially the same information as Figure~\ref{fig1}), but it illustrates its relevance due to the close connection to the observed death rate.

\bigskip

In terms of practical consequences, the estimated effect can be considered relevant and decreases monotonically. 

Under the assumption that relevant patient-specific risk factors have been sufficiently accounted for, this suggests that it may be worth to consider a volume-based intervention such as a minimum provider volume.

In addition, interventions that are not based on the volume (but may act on either $\boldsymbol{C}$ or $\boldsymbol{S}$ directly, see Section \ref{ssec:dag}) should be considered. Especially if the correlations induced by $\boldsymbol{C}$ are strong, a volume-based intervention may be less effective than the estimated volume-outcome relationship suggests. Moreover, the substantial uncertainty associated with the estimated volume effect may lead to an over- or underestimation of the impact of the intervention.

The strong volume-independent provider effects suggest that besides the focus on providers with small volumes, improvements also for providers with larger volumes should be aimed at.

Apart from assessing the effectiveness of possible interventions in terms of the considered outcome, decision makers must carefully weigh the expected benefit against potential negative side effects, e.g. health care shortage in certain regions.

\section{Discussion}\label{sec:discussion}

In this paper, we have presented, discussed and applied a versatile approach for the statistical analysis of volume-outcome associations in health care based on individual patient data.

The discussion of our DAG in Figure \ref{fig:DAG} (Section \ref{ssec:dag}) leads to two requirements for interpreting the estimated volume-outcome association as a causal effect instead of simply stating that there is an association. 
Fulfilling or verifying these requirements may certainly prove difficult as it essentially depends on information one cannot access directly. Furthermore, the DAG only incorporates possible interventions in a simplified way as they will usually affect the provider landscape as a whole rather than only a single provider. For example, if a provider cannot meet new regulations as part of an intervention, demand from other providers will increase. 

That touches the additional question of transportability\cite{PearlBareinboim2011:Transportability}: The relationship is usually estimated from observational data, and one has to ask how this relationship changes in the course of the intervention. Is it reasonable to assume that the `new' providers after the intervention are similar to those `old' providers in the fitting data that have a comparable volume? Predictions will tend to be more reliable if they concern rather moderate interventions (such as small modifications of minimum caseload requirements) that preserve a large fraction of larger hospitals.

Moreover, one has to take into account that the effect of any intervention will not be immediate, because it takes some time until the mechanisms behind the volume-outcome associations start to have an impact.

\bigskip

This work opens up a number of interesting directions for future developments.
There are statistical methods to estimate potential threshold values for the volume, e.g., based on break point models. Apart from that, one may employ the estimated volume-outcome association to quantify the potential effect of certain interventions: 
What is the expected number/proportion of adverse events which could be prevented through a suitable intervention?
However, note that the statistical analysis of the volume-outcome relationship alone does not suffice to answer the question if any proposed intervention is actually prudent in an administrative or clinical sense. For instance, the predicted benefit needs to be balanced against potential negative side-effects such as an increase of the travel distance from patients to providers\cite{heller2008auswirkungen, ChristiansenVrangbaek18:CentralizationDenmark}. 

Moreover, we focus on a logistic model (binary outcome) as that case is common in health care. 
However, clearly, central ingredients such as the smooth effect modelled through splines and random intercepts agree just as well with other outcome distributions (e.g. Gaussian or multinomial).

\bigskip
A volume-outcome analysis based on observational data will in many cases be the best way to generate a meaningful estimate of the volume effect and possible administrative decisions will therefore strongly depend on it. 
In this work, we proposed a flexible modelling approach that relies on a joint analysis of the effect of the patient-specific risk factors, the effect of provider volume and the effect of volume-independent provider characteristics. The approach puts particular emphasis on a flexible, smooth volume effect covering a wide range of possible monotonous or non-monotonous volume-outcome associations as well as the clustered structure of the data. We paved the way to go beyond reporting pure associations by stating conditions under which the volume effect can be interpreted causally.

%\begin{acks}
\paragraph{Acknowledgements}~\\
The authors thank Michael Höhle (IQTIG) for his general support and for providing helpful comments on the manuscript. Furthermore, the authors appreciate the collaboration with Günther Heller, Teresa Thomas and Janina Sternal (all at IQTIG) who gave feedback on the statistical methodology and provided expertise on the VLBW-data. 
The authors also wish to thank Helmut Küchenhoff (LMU Munich) for his valuable advice on the statistical approach. 
%\end{acks}

\newpage

\paragraph{Declaration of conflicting interests}~\\
The authors declare that there is no conflict of interest.

\appendix

\section{Construction of confidence intervals for odds ratios}\label{sec:or_est}

Asymptotic confidence intervals for the odds ratios $\mbox{OR}(\vol_1,\vol_2)$ can be obtained from the Delta method\cite{agresti2003categorical}.

The fitting procedure of the GAMM outputs both an estimate $\boldsymbol{\widehat{\theta}}$ of the parameter vector 
Let $\widetilde{\boldsymbol{X}}^\ast$ be the design matrix (including entries $s_{1:c}(\vol_i)$ for the spline basis) of two arbitrary patients which differ only in terms of $\vol$ such that the first row contains $s_{1:c}(\vol_1)$ and the second row contains $s_{1:c}(\vol_2)$. Then, by construction,

\begin{align*}g(\boldsymbol{\widehat{\theta}}):= \widehat{\mbox{OR}}(\vol_1,\vol_2) & = \exp\left((1,-1)\cdot \widetilde{\boldsymbol{X}}^\ast\cdot \boldsymbol{\widehat{\theta}}\right)\\
&= \exp\left(\widehat{\fvol}(\vol_1)-\widehat{\fvol}(\vol_2)\right) %\\
% &= \widehat{\mbox{OR}}(\vol_1,\vol_2),~~
\qquad\mbox{and so}\\
\nabla g(\boldsymbol{\widehat{\theta}}) &= (1,-1)\cdot \widetilde{\boldsymbol{X}}^\ast\cdot g(\boldsymbol{\widehat{\theta}}),
\end{align*}
the latter expression being a row vector.
The estimated standard error of $g(\boldsymbol{\widehat{\theta}})$ is
$$
\widehat{\sigma}_g = \nabla g(\boldsymbol{\widehat{\theta}})\cdot \widehat{\Sigma} \cdot \left(\nabla g(\boldsymbol{\widehat{\theta}})\right)^\top.
$$
The corresponding asymptotic 95\%-confidence-interval is given by
$$
\left[g(\boldsymbol{\widehat{\theta}}) - 2\widehat{\sigma}_g, g(\boldsymbol{\widehat{\theta}}) + 2\widehat{\sigma}_g\right].
$$

\section{Further simulation results}\label{sec:more_simulations}

This supplemental material contains further simulation results for different volume-outcome relationships and for different values of the standard deviation $\tau$ of the provider effects.  The setup is the same as explained in Section~\ref{sec:simulations}.

\subsection{Results for different volume-outcome relationships}

\begin{figure}
\centering
\includegraphics[width=0.9\textwidth]{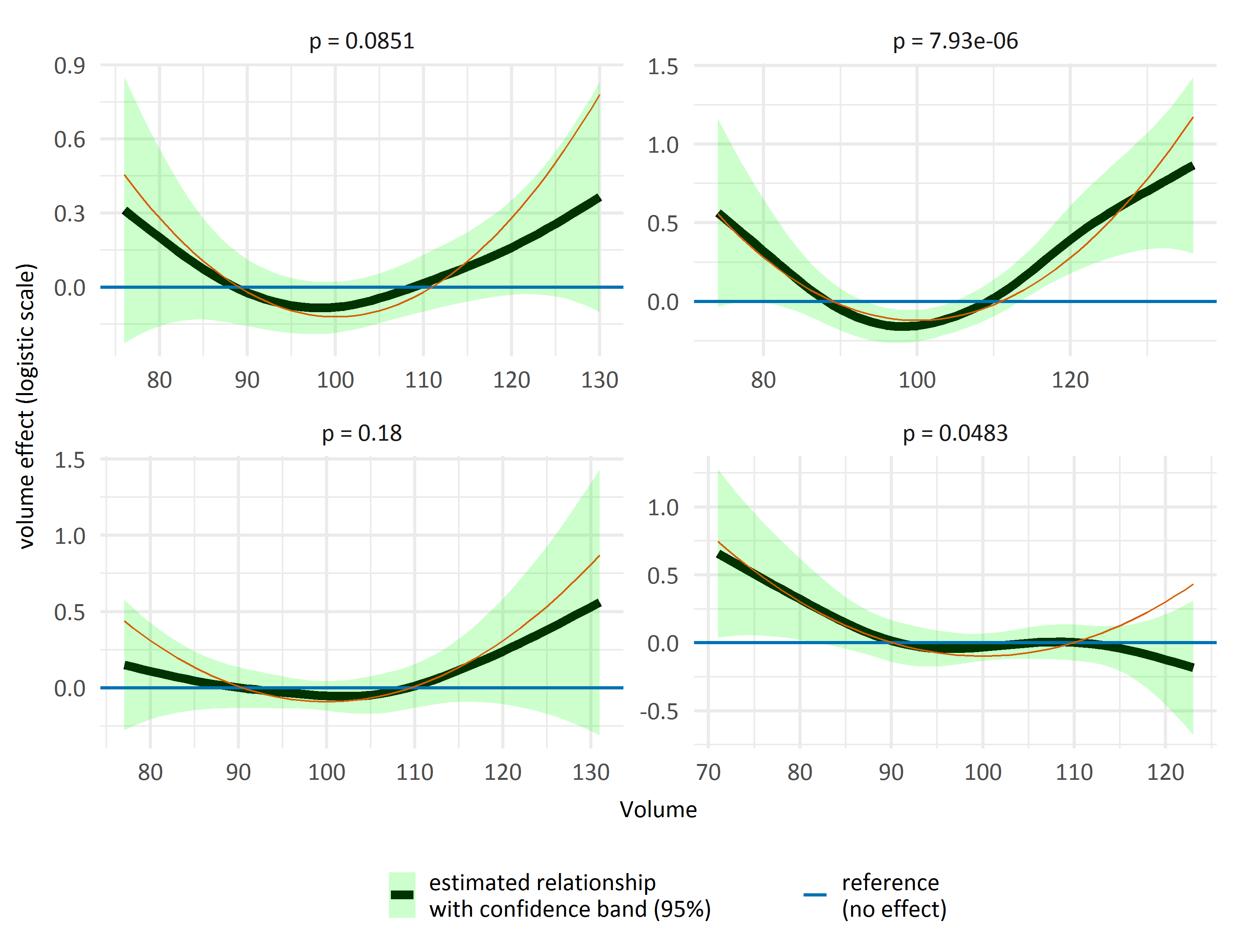}
\caption{The estimated volume effect $\widehat{\fvol}$ in four simulation runs with a U-shaped volume-outcome relationship, $I=200$ and $\tau=0.5$.  The true effect is plotted in red.\label{fig:simpredplot}}
\end{figure}

\begin{figure}
\centering
\includegraphics[width=0.9\textwidth]{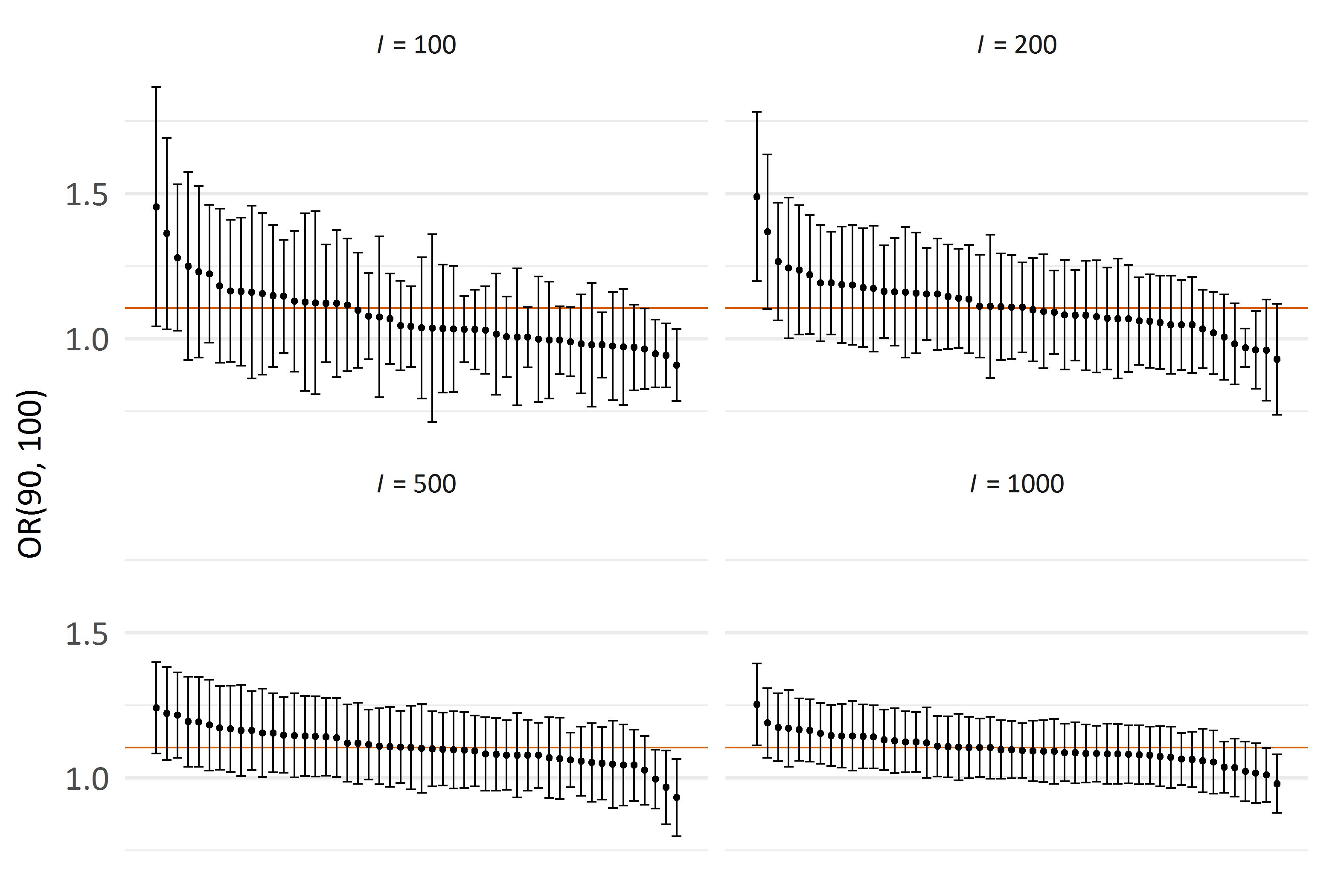}
\caption{Estimates of the odds ratio OR(90,100) for 200 simulation runs with different values of $I$ with a U-shaped volume-outcome relationship.\label{fig:ORplot_sel_v2}}
\end{figure}

Figure~\ref{fig:simpredplot} shows the estimate of the volume effect for four simulation runs with a U-shaped true volume effect, as in the main text.
Figure~\ref{fig:ORplot_sel_v2} shows estimates of OR(90,100) including confidence intervals (constructed as explained in Section \ref{sec:or_est}) for all simulation runs for four values of the number of hospitals.

\begin{figure}
\centering
a)\raisebox{-0.4\textwidth}{\includegraphics[width=0.45\textwidth]{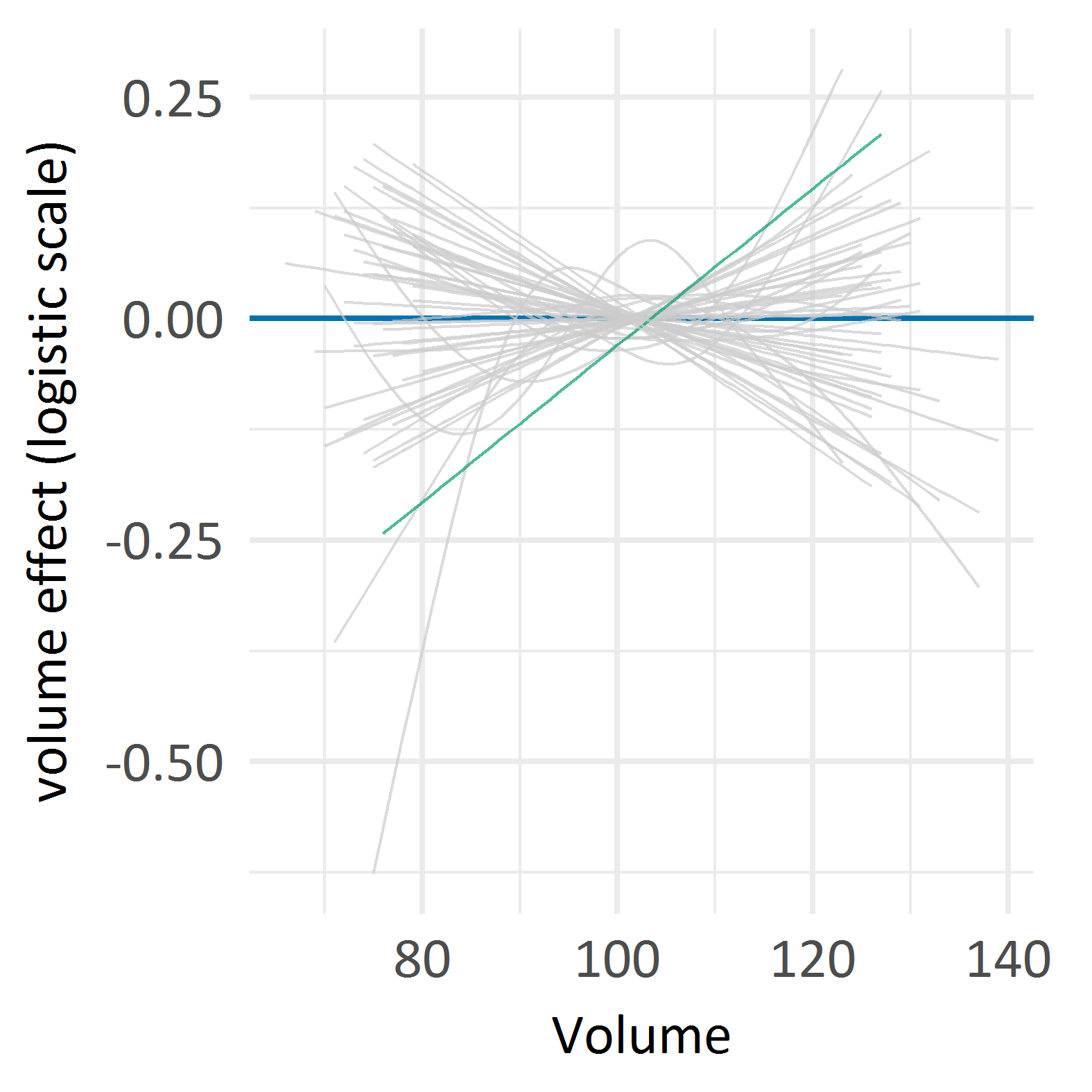}}
b)\raisebox{-0.4\textwidth}{\includegraphics[width=0.45\textwidth]{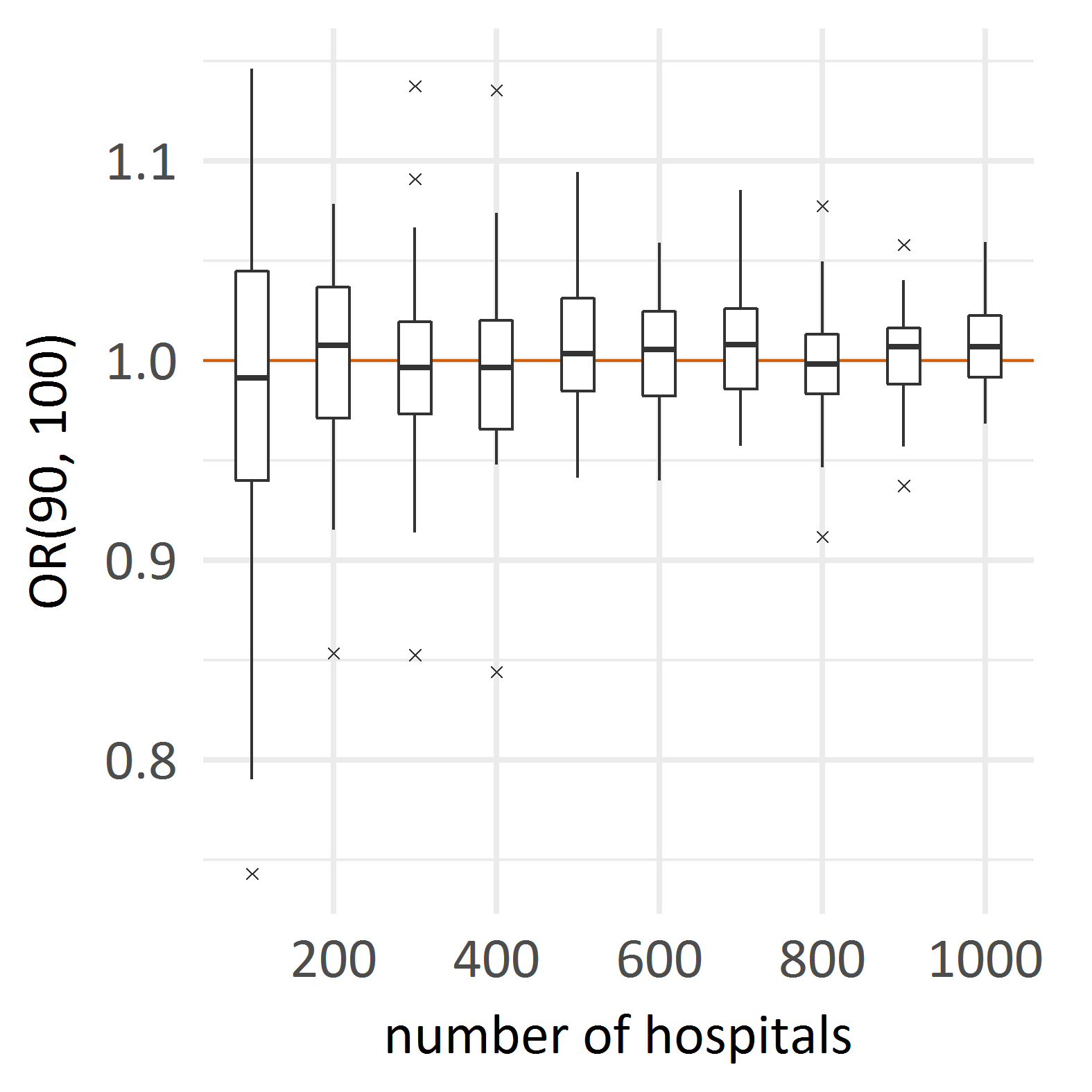}}
\caption{a)~The estimated volume effect $\widehat{\fvol}$ in 50 simulation runs without volume effect, $I=200$ and $\tau=0.5$. Curves are green if the $p$-value of the test for a volume effect is $\le 0.05$.\label{fig:simpredplotall_v0}
b)~Estimates of the odds ratio OR(90, 100) obtained in several simulation runs. For each number of hospitals, 50 runs were made.
The true value is plotted in red.\label{fig:simORplot_v0}}
\end{figure}

\begin{figure}
\centering
\includegraphics[width=0.9\textwidth]{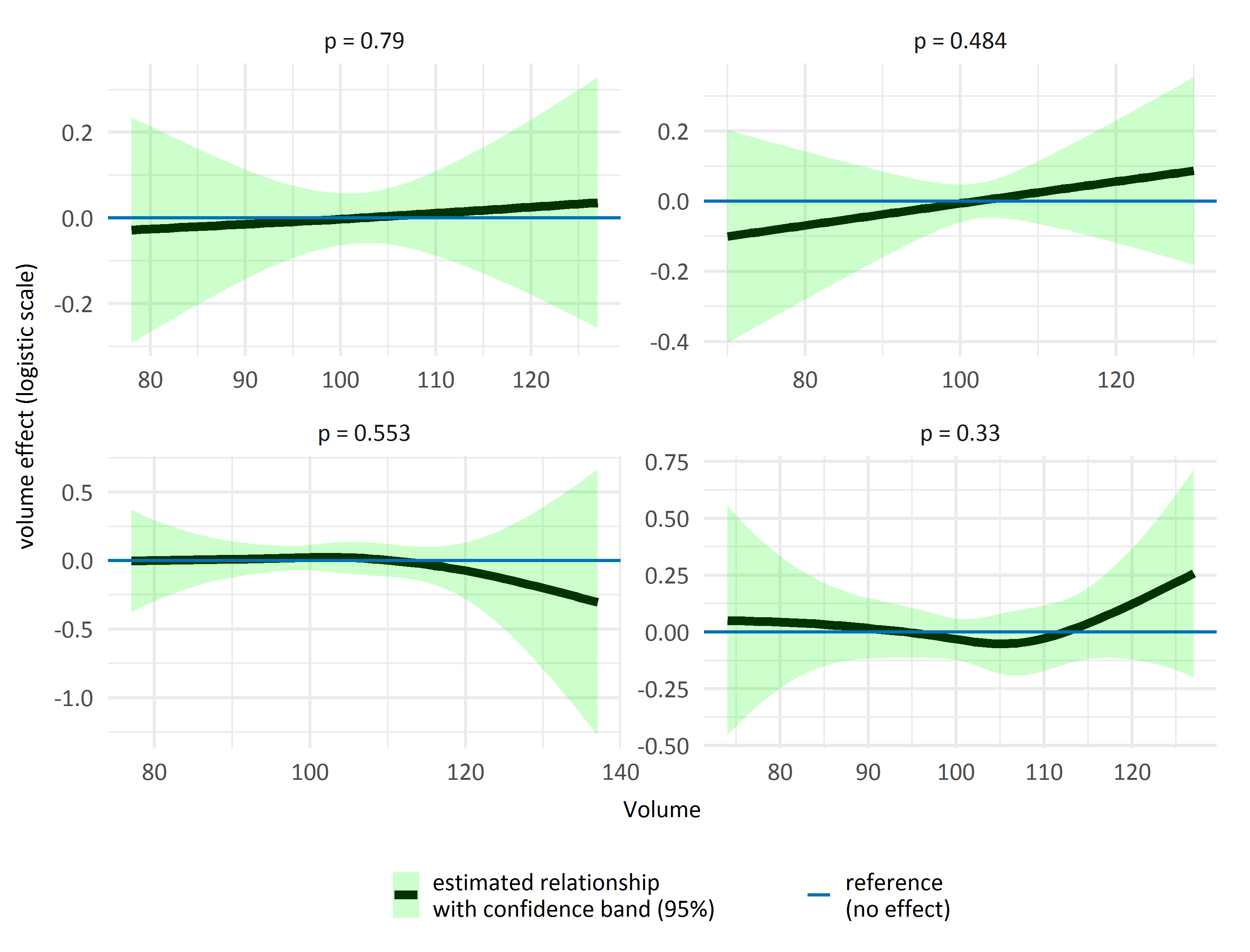}
\caption{The estimated volume effect $\widehat{\fvol}$ in four runs of our simulation study with no volume-outcome relationship, $I=200$ and $\tau=0.5$.\label{fig:simpredplot_v0}}
\end{figure}

\begin{figure}
\centering
a)\raisebox{-0.4\textwidth}{\includegraphics[width=0.45\textwidth]{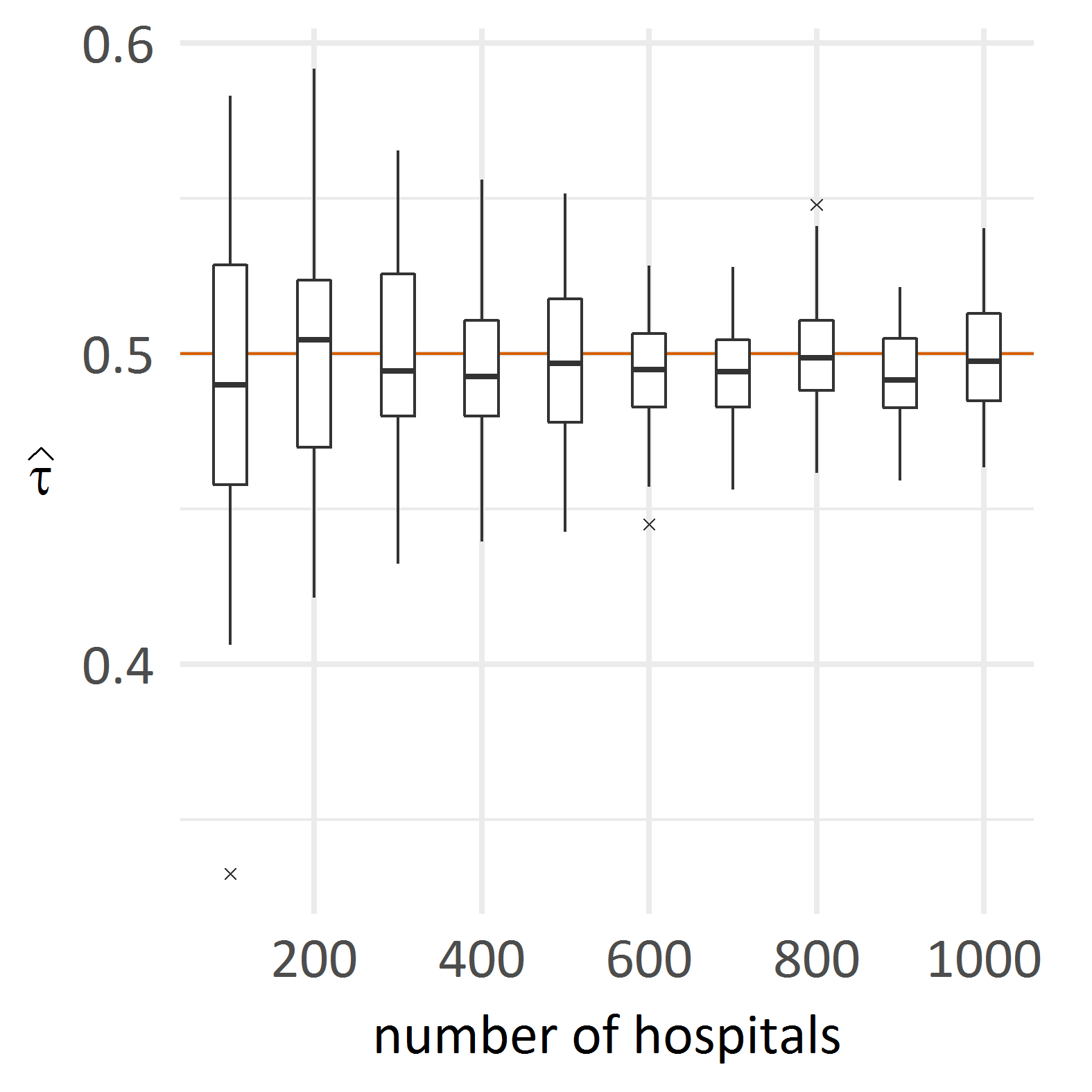}}
b)\raisebox{-0.4\textwidth}{\includegraphics[width=0.45\textwidth]{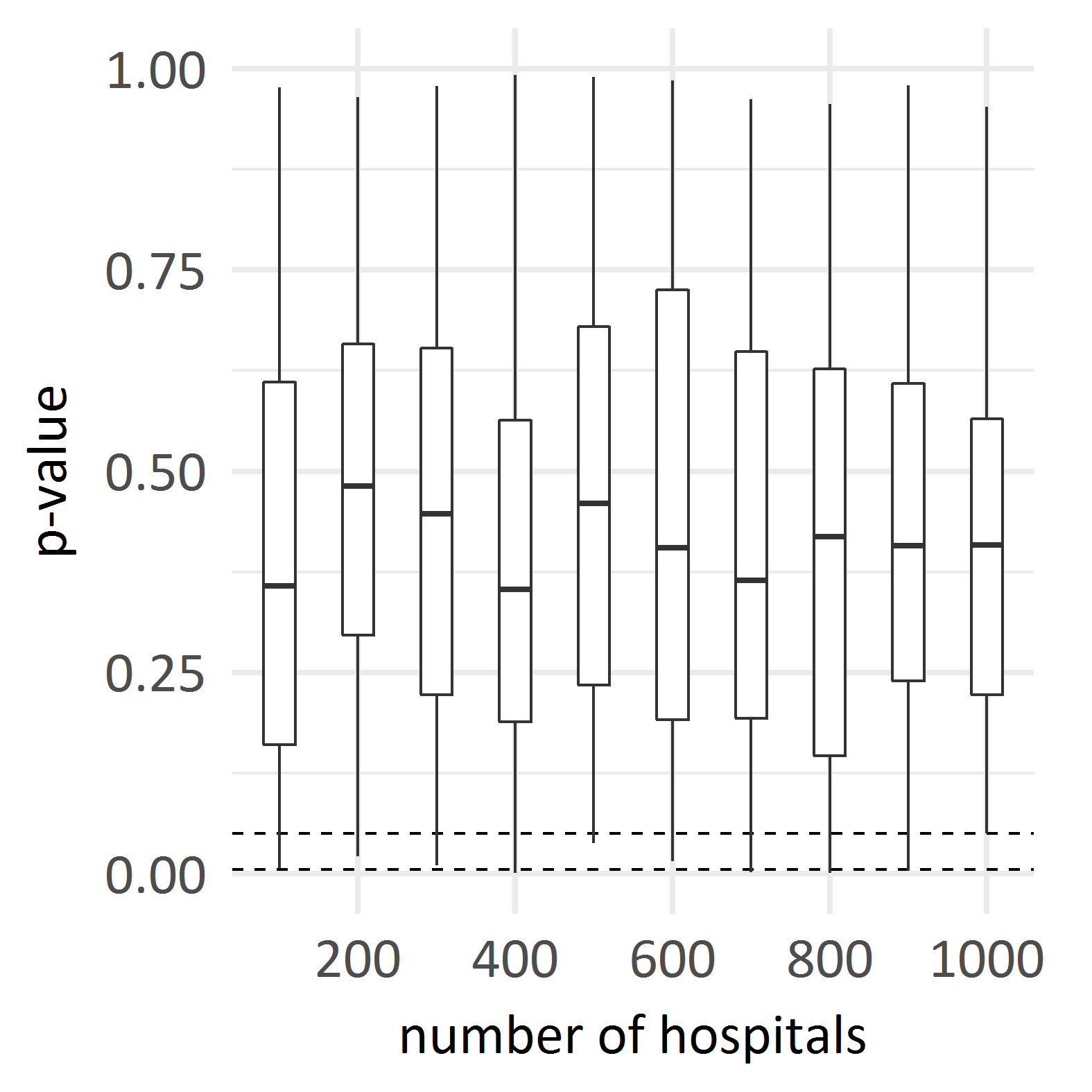}}
\caption{
  The estimate $\widehat\tau$ (a) and the $p$-value of the test for a volume effect (b) of several simulation runs with no volume-outcome relationship.  \label{fig:simtau_v0}
  For each number of hospitals, 50 runs were made.  In b), horizontal lines mark the significance values of 0.05 and 0.005.\label{fig:simps_v0}}
\end{figure}

Figures~\ref{fig:simpredplotall_v0}, \ref{fig:simpredplot_v0} and~\ref{fig:simtau_v0} show estimates of the volume effect, estimates of $\tau$ and $p$-values of the tests for volume effects in the case of no volume-outcome relationship.
In most runs, the estimated volume effect is small.
As expected, the $p$-values are spread out evenly over the unit interval.
The estimated volume effect is significant at significance level 0.05 in 25 of the 500 runs, which coincidently corresponds to a fraction of precisely 5 percent. % yeah.

\begin{figure}
\centering
\includegraphics[width=0.9\textwidth]{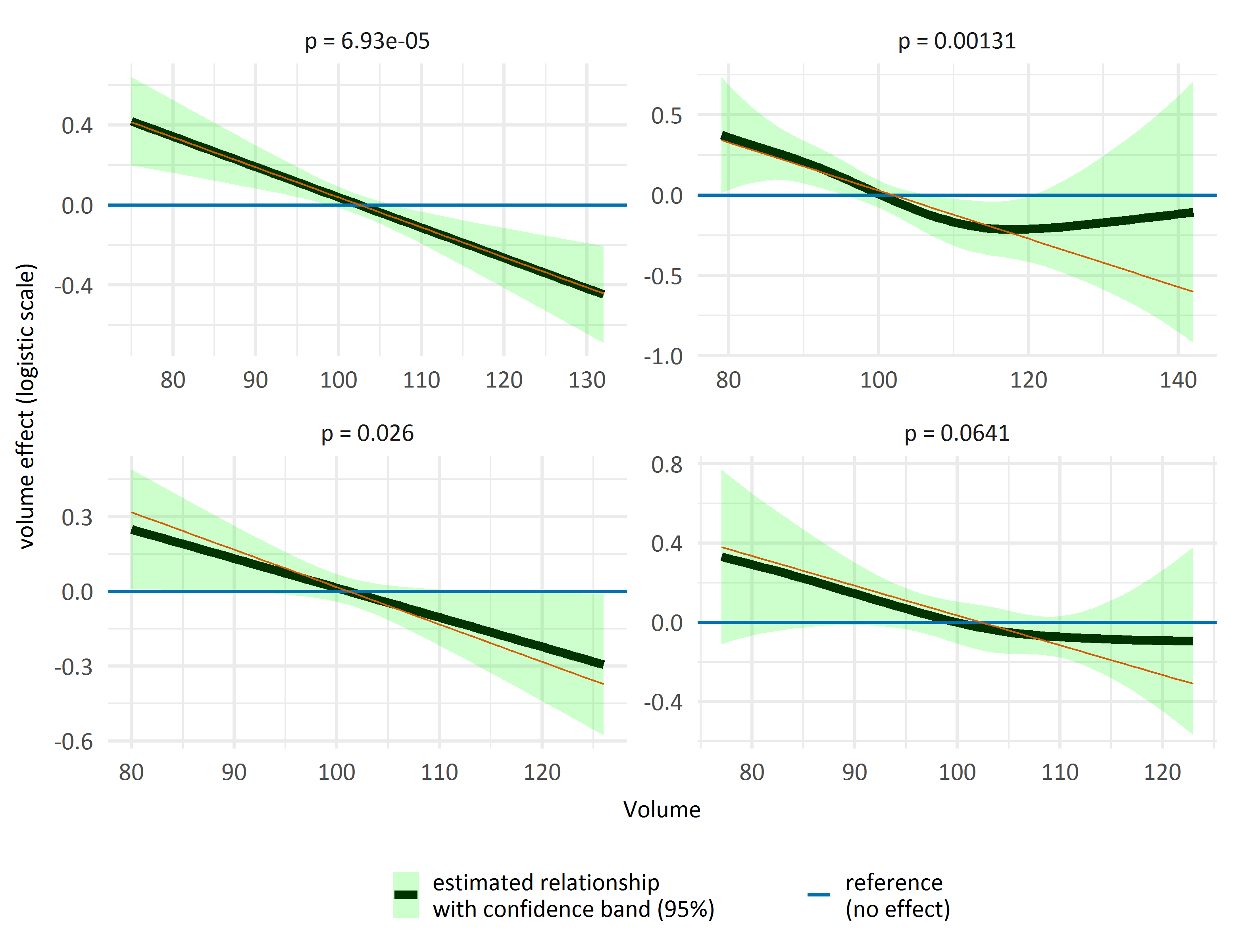}
\caption{The estimated volume effect $\widehat{\fvol}$ in four runs of our simulation study with a linear volume-outcome relationship, $I=200$ and $\tau=0.5$.  The true effect is plotted in red.\label{fig:simpredplot_v1}}
\end{figure}

\begin{figure}
\centering
a)\raisebox{-0.4\textwidth}{\includegraphics[width=0.45\textwidth]{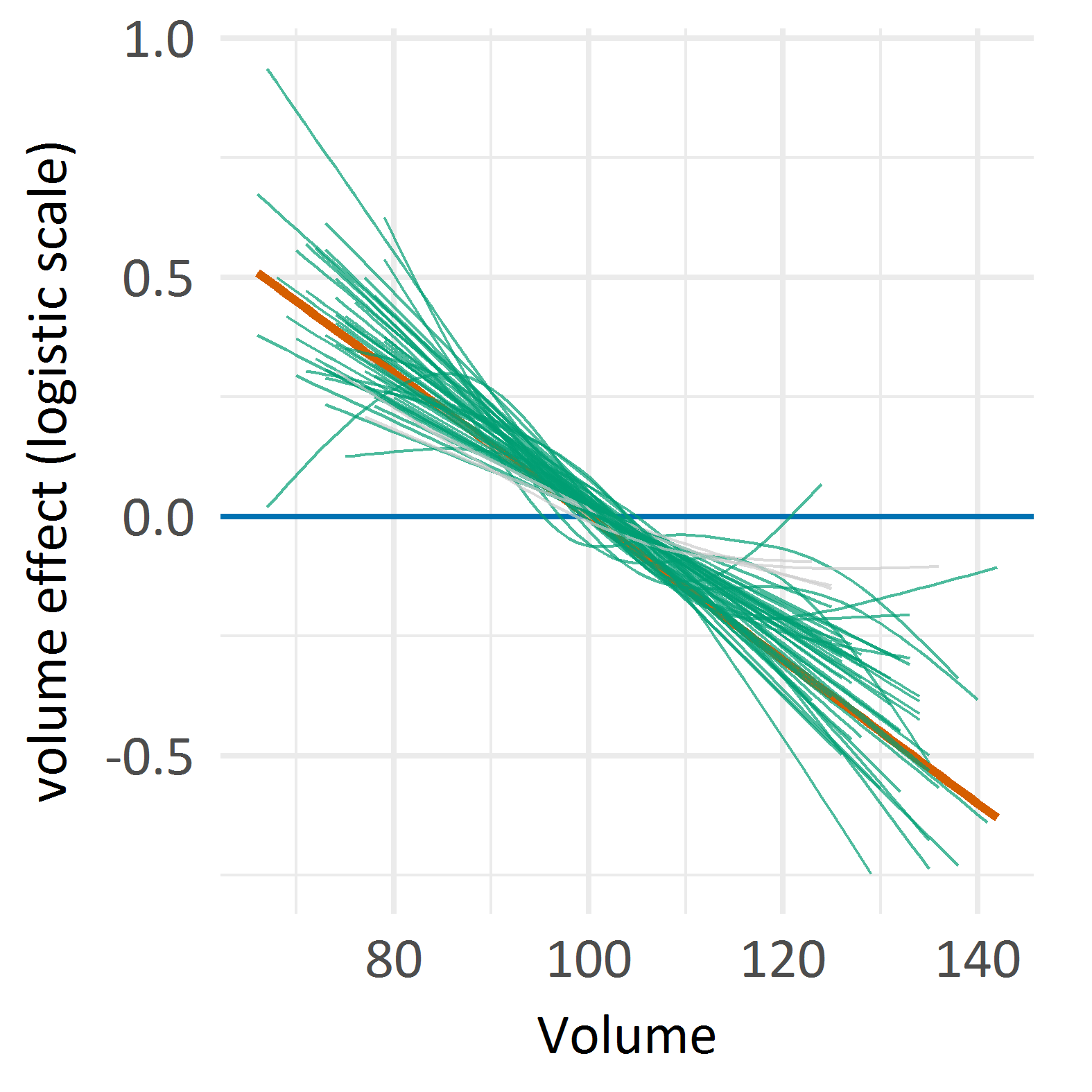}}
b)\raisebox{-0.4\textwidth}{\includegraphics[width=0.45\textwidth]{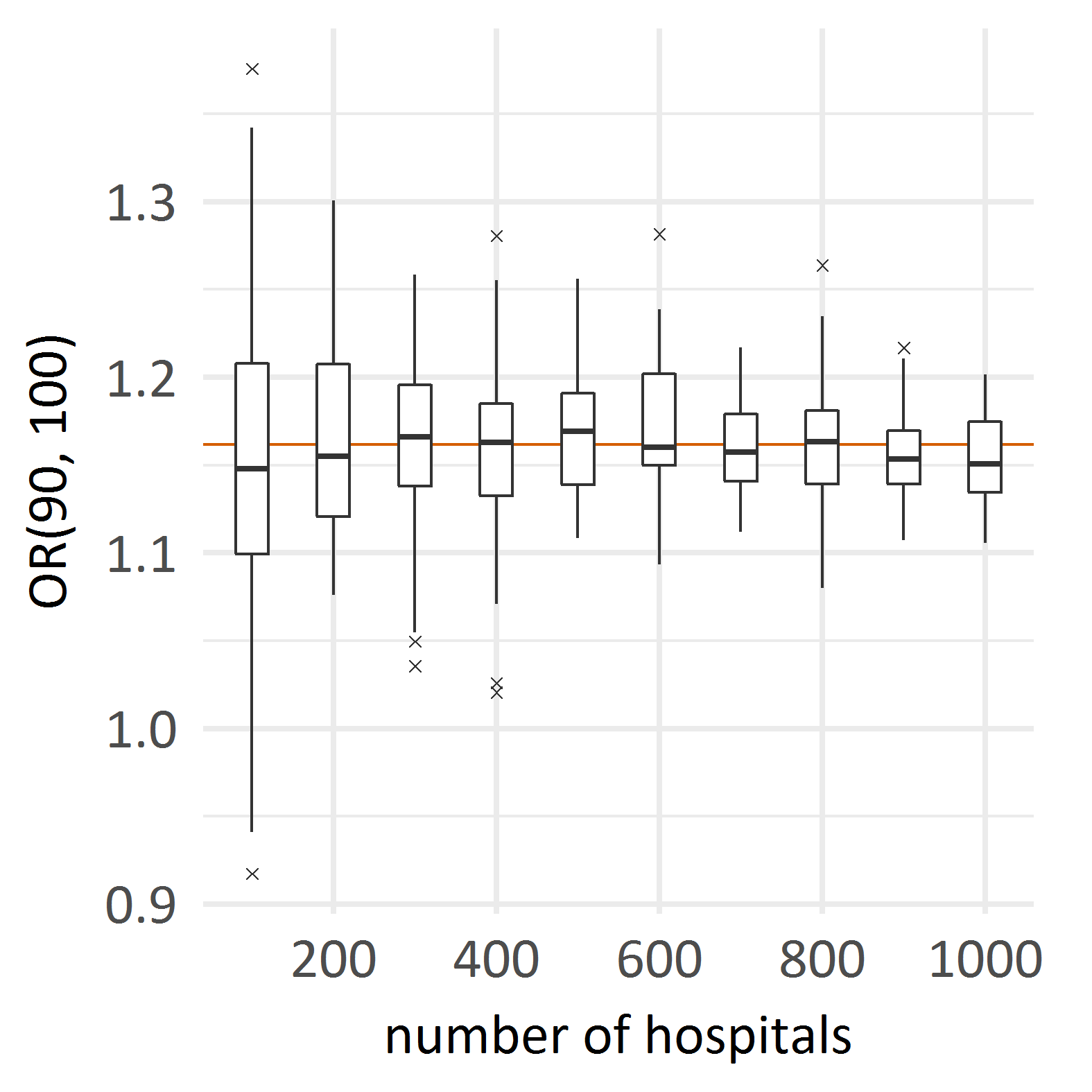}}
\caption{a)~The estimated volume effect $\widehat{\fvol}$ in 50 simulation runs with a linear volume-outcome relationship, $I=200$ and $\tau=0.5$.  The true effect is plotted in red. Curves are green if the $p$-value of the test for a volume effect is $\le 0.05$.\label{fig:simpredplotall_v1}
b)~Estimates of the odds ratio OR(90, 100) obtained in several simulation runs. For each number of hospitals, 50 runs were made.
The true value is plotted in red.\label{fig:simORplot_v1}}
\end{figure}

\begin{figure}
\centering
a)\raisebox{-0.4\textwidth}{\includegraphics[width=0.45\textwidth]{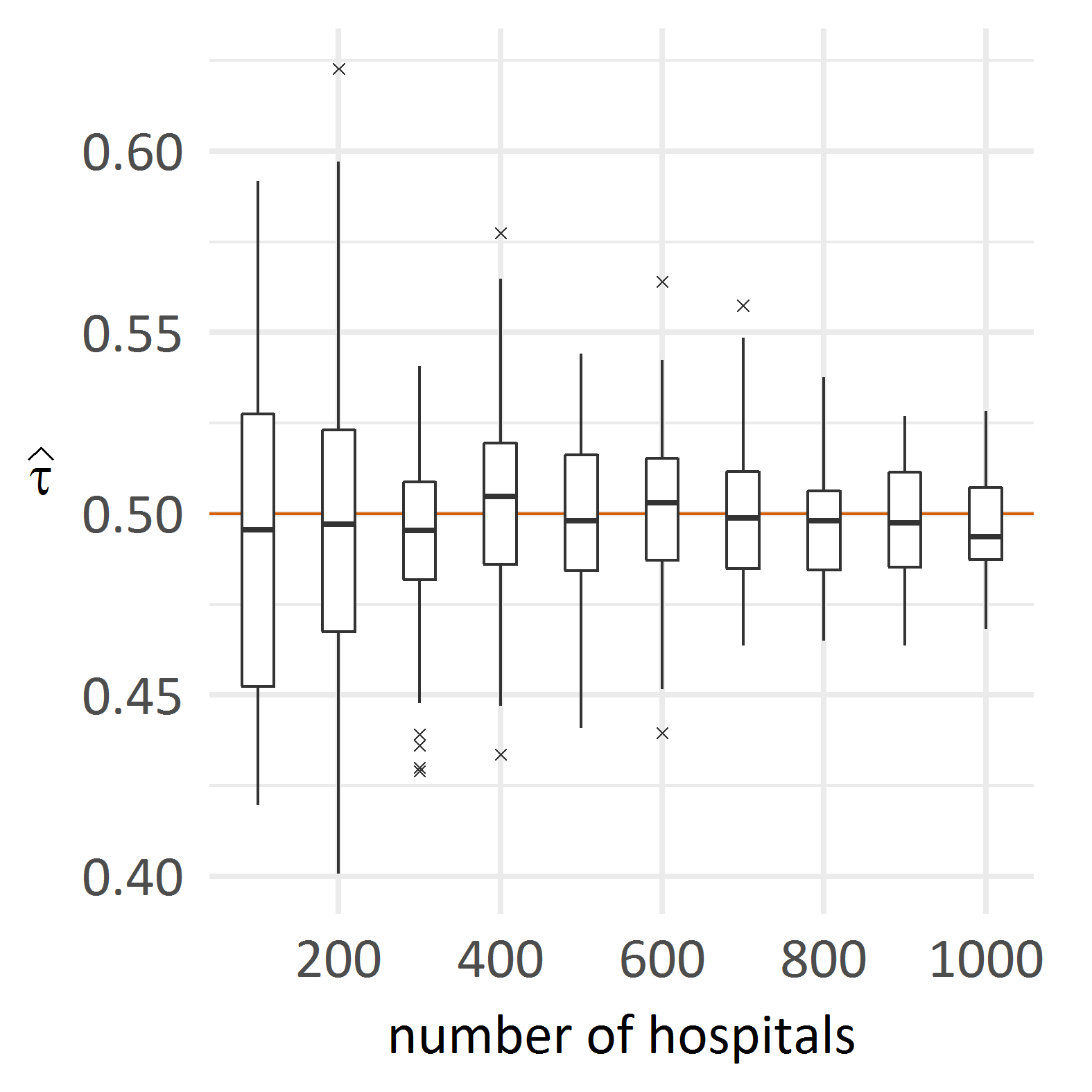}}
b)\raisebox{-0.4\textwidth}{\includegraphics[width=0.45\textwidth]{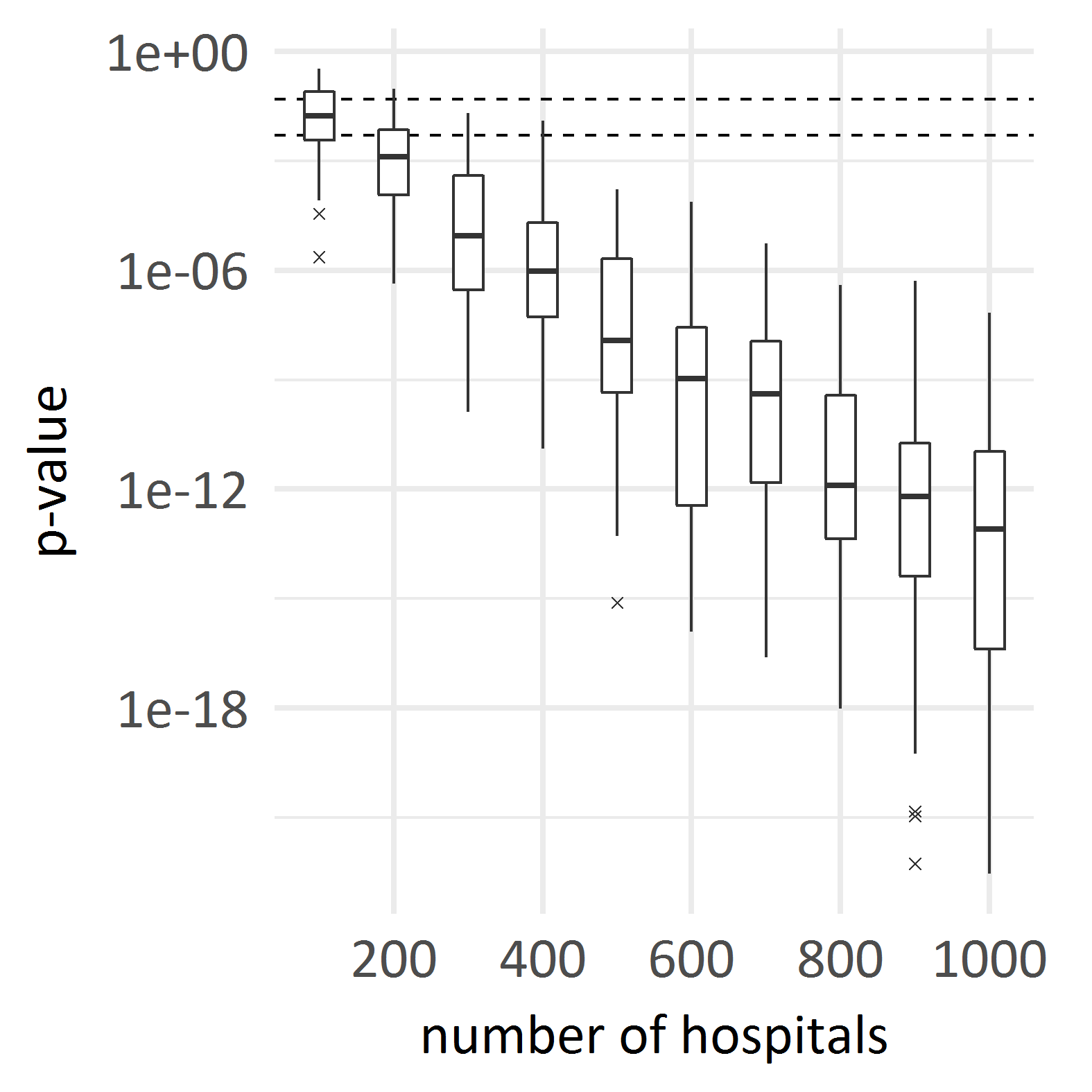}}
\caption{
  The estimate $\widehat\tau$ (a) and the $p$-value of the test for a volume effect (b) of several simulation runs with a linear volume-outcome relationship.  \label{fig:simtau_v1}
  For each number of hospitals, 50 runs were made.  In b), dashed horizontal lines mark the significance values of 0.05 and 0.005.\label{fig:simps_v1}}
\end{figure}

Figures~\ref{fig:simpredplot_v1}, \ref{fig:simORplot_v1} and~\ref{fig:simtau_v1} concern the case of a linear volume-outcome relationship: $\fvol(n) = \frac{3}{100}(100 - n)$.
Again, the estimated volume effects lie reasonably close to the ground truth.

As in the main text, in the two settings presented in this supplement, the provider effect was always significant ($p$-value less than $10^{-9}$ in all simulation runs).

\subsection{Results for different values of \texorpdfstring{$\tau$}{tau}}

\begin{figure}
\centering
a)\raisebox{-0.4\textwidth}{\includegraphics[width=0.45\textwidth]{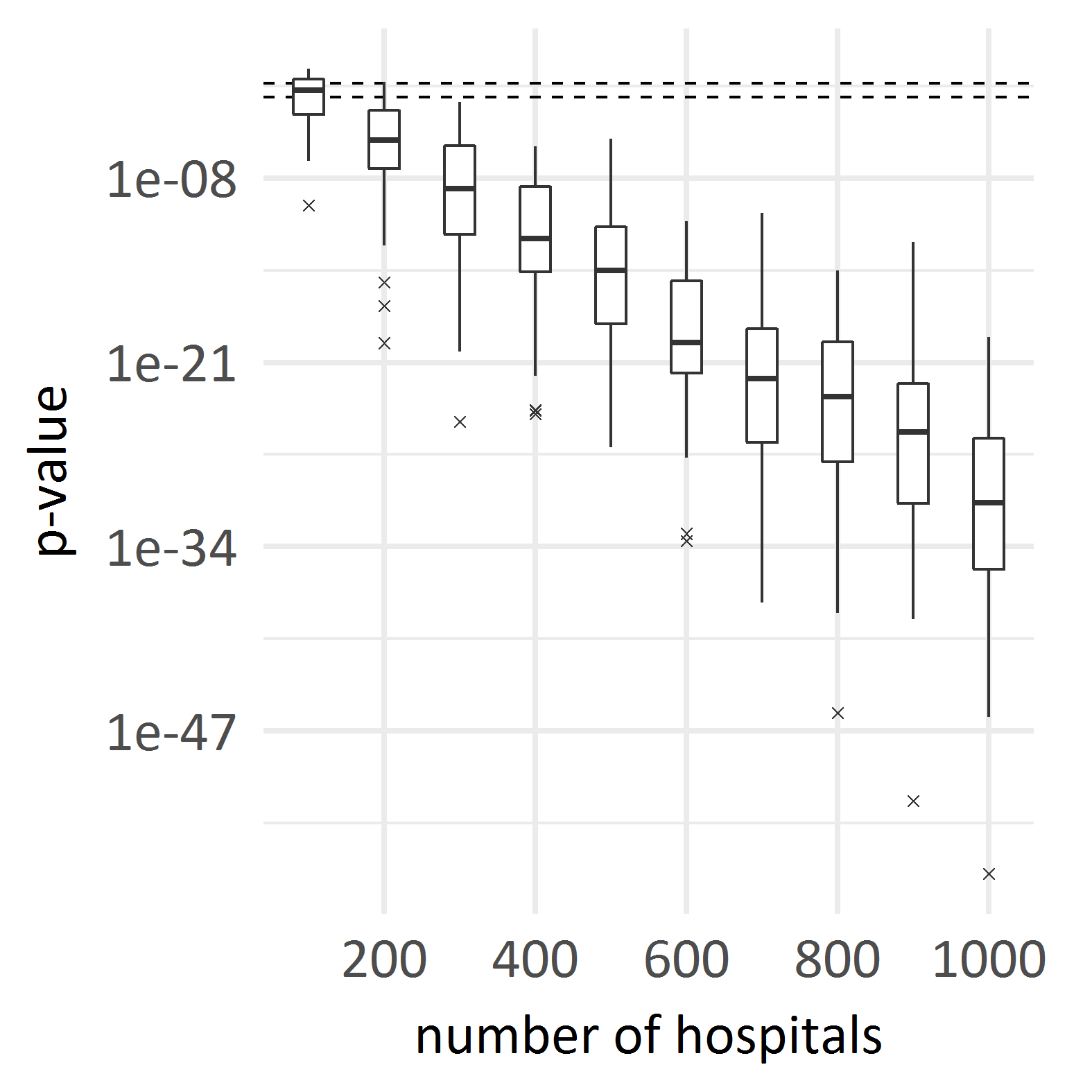}}
b)\raisebox{-0.4\textwidth}{\includegraphics[width=0.45\textwidth]{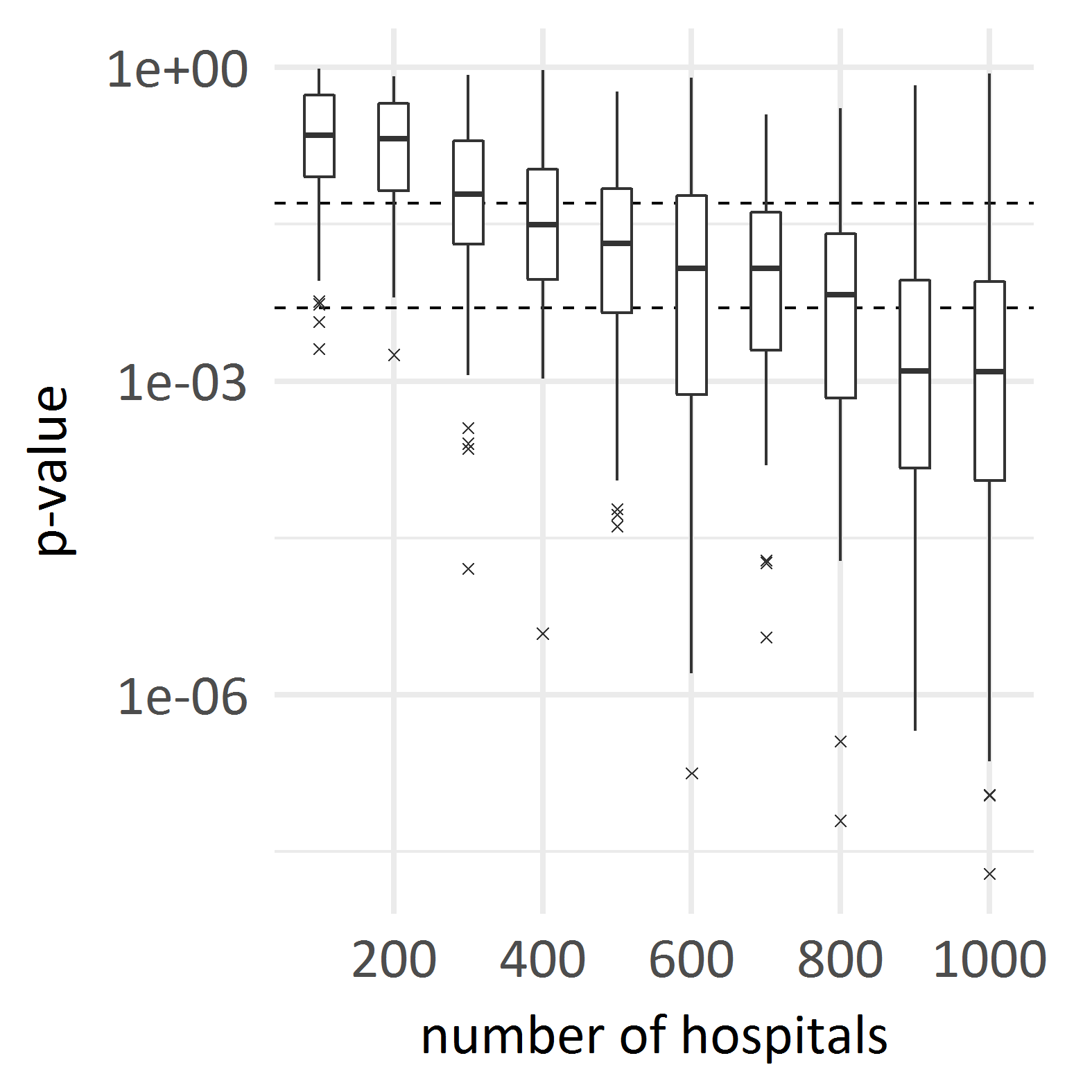}}
\caption{
  The $p$-value of the test for a volume effect of several simulation runs with a U-shaped volume-outcome relationship and with $\tau = 0.25$ (a) and $\tau = 1$ (b).
  For each number of hospitals, 50 runs were made.  Horizontal line mark the significance values of 0.05 and 0.005.\label{fig:sim_taus}}
\end{figure}

To study the impact of $\tau$, all simulations were repeated for $\tau = 0.25$ and $\tau = 1.0$.
The results for other values of $\tau$ are similar. However, when $\tau$ becomes larger, it becomes more difficult to reliably detect a volume effect. To illustrate this, Figure~\ref{fig:sim_taus} shows the p-values of the volume effects for simulation runs with a U-shaped volume outcome relationship and all other parameters as in the main text, but with different values of $\tau$.

\section{Volume variability in the VLBW data}\label{sec:NICU_volume_plot}

\begin{figure}
\centering
\includegraphics[width=25pc,height=15pc]{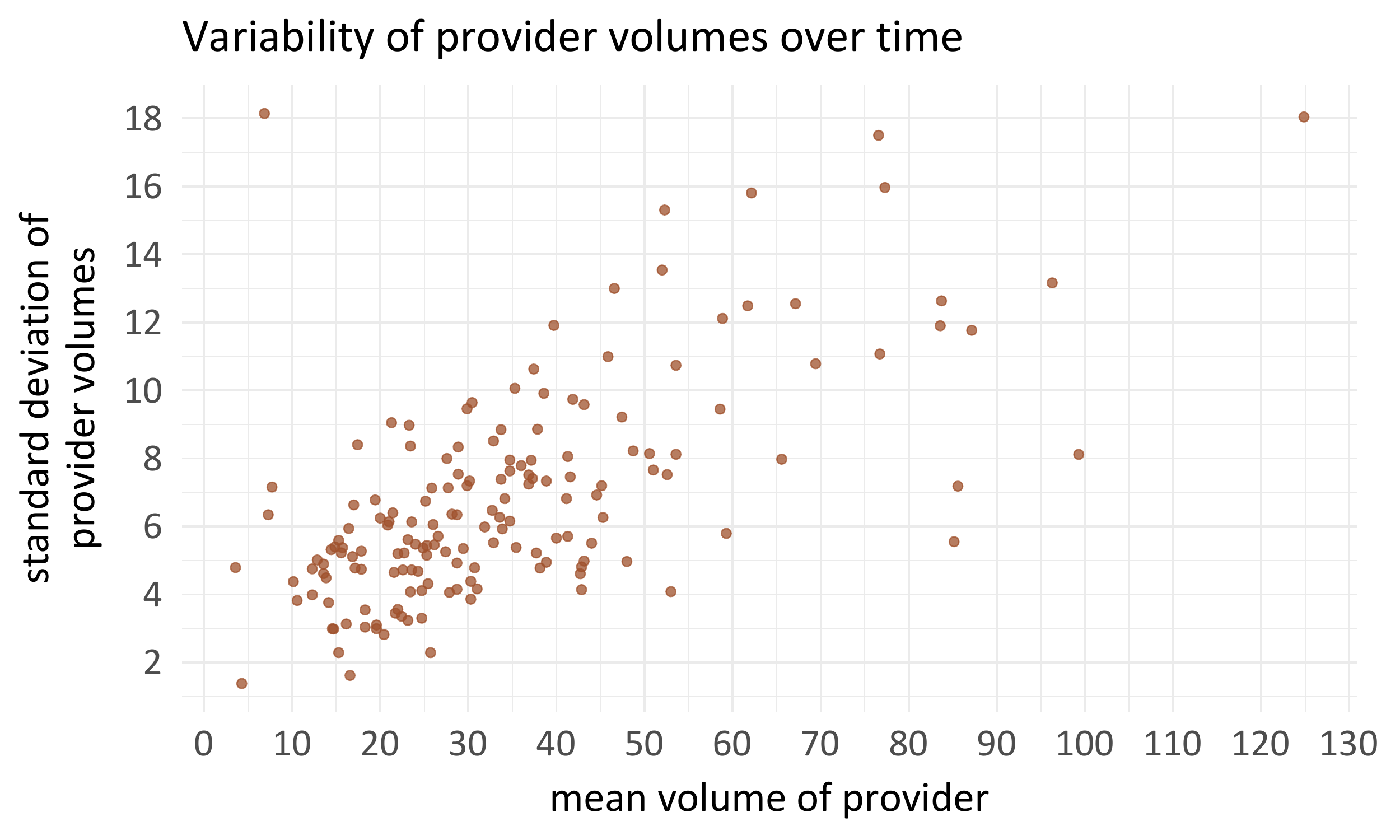}
\caption{Illustration of the oberserved variability of the provider volumes in the VLBW data across 7 years.\label{fig:NICU_volume_plot}}
\end{figure}

Figure \ref{fig:NICU_volume_plot} displays the mean of the volumes $\widetilde{v}_i^{2012}, \widetilde{v}_i^{2013}, \ldots, \widetilde{v}_i^{2018}$ against the corresponding standard deviation for each provider $i$.  In case a provider did not treat VLBW over the whole period, mean and standard deviation are based on the existing data. 
Indeed, many providers---small and large ones---exhibit considerable volume fluctuations, which supports our decision to construct the more stable proxy from \eqref{eq:stable_proxy}. For instance, taking a closer look at the provider on the far right with mean volume of approximately $125$, the values range from $104$ to $148$. 

\FloatBarrier

\bibliographystyle{abbrv}
\bibliography{modelling_volout_relationships}

\end{document}